\documentclass[fleqn,usenatbib]{mnras}
\pdfoutput=1

\usepackage{newtxtext}

\usepackage[T1]{fontenc}
\usepackage{ae,aecompl}

\usepackage{etoolbox}
\makeatletter
\patchcmd\@combinedblfloats{\box\@outputbox}{\unvbox\@outputbox}{}{%
   \errmessage{\noexpand\@combinedblfloats could not be patched}%
}%
\makeatother

\usepackage{graphicx}	
\usepackage{amsmath}	
\usepackage{amssymb}	
\usepackage{longtable,lscape}
\usepackage{soul}

\newcommand{\tco}{$^{13}$CO}
\newcommand{\cdo}{C$^{18}$O}
\newcommand{\kms}{\mbox{km~s$^{-1}$}}
\newcommand{\Msun}{\mbox{$\text{M}_{\sun}$}}
\newcommand{\tyso}{\mbox{HH\ 900} YSO} 
\usepackage[dvipsnames]{xcolor}

\title[ALMA looks into a tadpole]{Illuminating a tadpole's metamorphosis II: observing the on-going transformation with ALMA}

\author[M. Reiter et al.]{Megan Reiter,$^{1}$\thanks{E-mail: megan.reiter@stfc.ac.uk (MR)}
Andr{\'e}s E. Guzm{\'a}n,$^{2}$
Thomas J. Haworth,$^{3}$
Pamela D. Klaassen,$^{1}$
\newauthor Anna F. McLeod,$^{4,5}$ 
Guido Garay,$^{6}$
Joseph C. Mottram$^{7}$
\\
$^{1}$UK Astronomy Technology Centre, Blackford Hill, Edinburgh, EH9 3HJ, UK \\
$^{2}$National Astronomical Observatory of Japan, 2-21-1 Osawa, Mitaka, Tokyo 181-8588, Japan \\
$^{3}$Astronomy Unit, School of Physics and Astronomy, Queen Mary University of London, London E1 4NS, UK \\
$^{4}$Department of Astronomy, University of California Berkeley, Berkeley, CA 94720, USA\\
$^{5}$Department of Physics and Astronomy, Texas Tech University, PO Box 41051, Lubbock, TX 79409, USA\\
$^{6}$Departamento de Astronom\'{i}a, Universidad de Chile, Camino el Observatorio 1515, Las Condes, Santiago, Chile \\
$^{7}$Max Planck Institute for Astronomy, K\"onigstuhl 17, 69117 Heidelberg, Germany
}

\date{Accepted XXX. Received YYY; in original form ZZZ}

\pubyear{2019}

\begin{document}
\label{firstpage}
\pagerange{\pageref{firstpage}--\pageref{lastpage}}
\maketitle

\begin{abstract}

We present new Atacama Large Millimeter/submillimeter Array (ALMA) observations of the tadpole, a small globule in the Carina Nebula that hosts the HH~900 jet+outflow system. 
Our data include $^{12}$CO, $^{13}$CO, C$^{18}$O J=2-1, $^{13}$CO, C$^{18}$O J=3-2, and serendipitous detections of DCN J=3-2 and CS J=7-6.  
With angular resolution comparable to the \emph{Hubble Space Telescope (HST)}, our data reveal for the first time the bipolar molecular outflow in CO, seen only inside the globule, that is launched from the previously unseen jet-driving protostar (the \tyso). 
The biconical morphology joins smoothly with the externally irradiated outflow seen in ionized gas tracers outside the globule, tracing the overall morphology of a jet-driven molecular outflow. 
Continuum emission at the location of the \tyso\ appears to be slightly flattened perpendicular to outflow axis. 
Model fits to the continuum have a best-fit spectral index of $\sim 2$, suggesting cold dust and the onset of grain growth. 
In position-velocity space, $^{13}$CO and C$^{18}$O gas kinematics trace a C-shaped morphology, similar to infall profiles seen in other sources, although the global dynamical behaviour of the gas remains unclear. 
Line profiles of the CO isotopologues display features consistent with externally heated gas. 
We estimate a globule mass of $\sim 1.9$~M$_{\odot}$, indicating a remaining lifetime of $\sim 4$~Myr, assuming a constant photoevaporation rate. 
This long globule lifetime will shield the disk from external irradiation perhaps prolonging its life and enabling 
planet formation in regions where disks are typically rapidly destroyed. 
\end{abstract}




\section{Introduction}

Ionizing radiation permeates high-mass star-forming regions, sculpting the natal cloud and excavating newly born stars.  
On large scales, ionizing radiation clears low-density gas, contributing to the ultimate destruction of the cloud and possibly resupplying turbulence \citep[e.g.,][]{gritschneder2009,gritschneder2010,dale2011,dale2013,tremblin2013,walch2013,boneberg2015,dale2017}. 
Feedback also affects the much smaller scales of individual stars and their circumstellar (planet-forming) disks \citep[e.g.,][]{mann2014,winter2018,nicholson2019}.

In between these two extremes are small ($r<1$~pc), bright-rimmed clouds often seen in and around H~{\sc ii} regions \citep[e.g.,][]{smith2003,gahm2007,wright2012}. 
Some appear to harbor nascent protostars \citep[e.g.,][]{mccaughrean2002,sahai2012,reiter2015_hh900}, while others appear to have resisted collapse \citep[e.g.,][]{smith2004_finger,gahm2013,haikala2017}. 
Mass estimates for these globules range from planetary to stellar masses \citep{gahm2007,sahai2012,gahm2013,haikala2017}. 
Several theoretical models have explored how external irradiation affects the globules, including whether it may stimulate collapse through radiatively-driven implosion  \citep[RDI; e.g.,][]{bertoldi1989,lefloch1994,kessel-deynet2003,miao2009,bisbas2011,haworth2013}. 
If stimulated to collapse, small globules may contribute significantly to the low-mass end of the initial mass function (IMF).

Small, opaque globules are typically identified in images where they are seen in silhouette against the bright background of the H~{\sc ii} region. 
Most studies use narrowband optical images as these provide higher angular resolution than single-dish observations at the long wavelengths that probe cold, molecular gas \citep{bok1948,pottasch1956,pottasch1958,dyson1968,herbig1974,schneps1980,reipurth1983,gahm2007,wright2012,grenman2014}. 
More recent efforts have targeted some larger globules for millimeter observations to measure molecular gas masses and radial velocities. 
Beamsizes from these single-dish studies 
tend to be significantly larger than the globules themselves, so bulk properties are inferred from line ratios and profiles  \citep[e.g.,][]{sahai2012,gahm2013,haikala2017}. 
Determining how feedback affects the fate of small globules requires spatially and spectrally resolved observations of the structure and kinematics of the cold molecular gas.

\begin{figure*}
  \centering
    \includegraphics[width=\textwidth, trim=5mm 20mm 25mm 10mm]{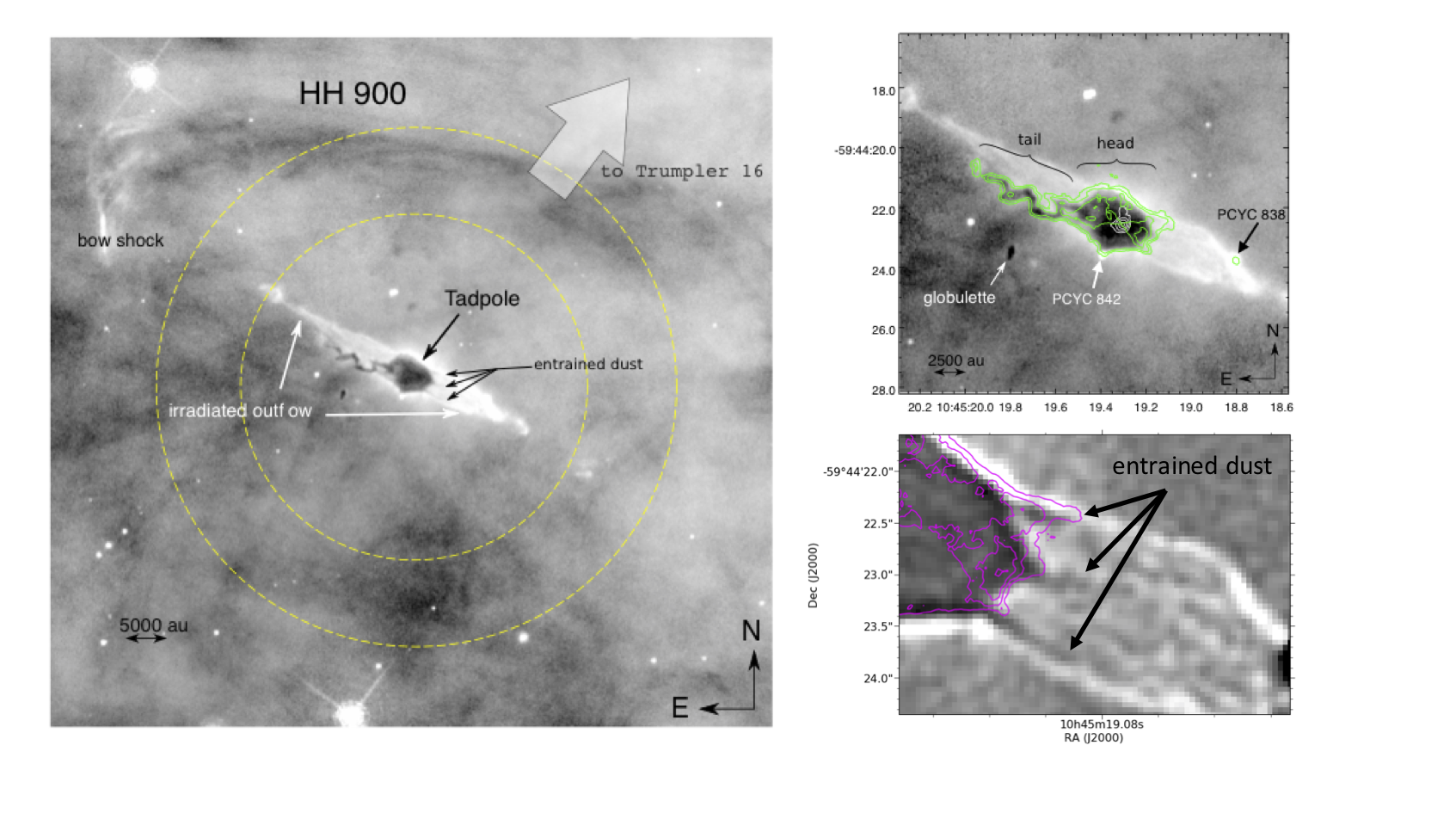}%
    \caption{ACS/HST H$\alpha$ image of the tadpole \citep{reiter2015_hh900}. \textit{Left:} Field of the tadpole. Dashed circles indicate the FWHM of Band 6 and 7 primary beams. \textit{Right:} White contours show ALMA continuum emission at 233 GHz (levels: 0.52, 0.78, 1.3, and 2.1 mJy beam$^{-1}$). Green contours show CO J=2-1 integrated over [-57, -11]~\kms. Intermediate contours display the bipolar outflow pattern of emission detected toward the tadpole (levels: 20, 50, 120, and 180 K \kms). \textit{Bottom Right:} A zoom on the globule with an unsharp mask applied to emphasize the three dust streamers (labelled). Magenta contours are the CO J=2-1 integrated intensity, as above.
      \label{fig:landscape}}
\end{figure*}

The subject of the present study is a small globule in the Carina Nebula, which we colloquially refer to as the tadpole (see Figure~\ref{fig:landscape}). 
Multiple O-type stars in the nearby cluster Tr16 illuminate the globule and the peculiar HH~900 jet+outflow system that emerges from it \citep{smith2010}. 
With a jet dynamical age of $\sim 2200$~yr \citep{reiter2015_hh900}, HH~900 is one of the youngest jets in Carina. 
Derived jet kinematics require a driving source embedded in the small opaque globule, but previous observations provided no evidence for a protostar inside the tadpole. 
Even at shorter wavelengths (i.e.\ 3.6~$\mu$m with \emph{Spitzer}), the angular resolution is comparable to the size of the globule, making it difficult to distinguish between emission from a embedded source and the two protostars that lie just outside the globule (see Figure~\ref{fig:landscape}).
Confusion only worsens toward longer wavelengths (with e.g., \emph{Herschel}) where young (Class~0) protostars emit the majority of their radiation.
At the southern declination of Carina, only the Atacama Large Millimeter/submillimeter Array (ALMA) provides the requisite angular resolution to detect an embedded protostar and the structure and kinematics of the surrounding globule.

A word on terminology: 
We use ``outflow'' to describe wide-angle flows that may be ambient material entrained by the jet or may originate from the disk wind \citep[see, e.g.,][]{klaassen2015}. 
These typically trace slower emission, with velocities $\sim 10$s~\kms, and are often observed in molecular gas tracers like CO, but in highly irradiated regions like Carina, may also be seen in ionized gas tracers like H$\alpha$ \citep[see, e.g.,][]{reiter2015_hh900,reiter2015_hh666}. 
This is different from the ``jet'' which we use to refer to the fast ($\sim 100$~km~s$^{-1}$), collimated (opening angles $<10^{\circ}$) stream of emission most often seen in the optical and near-IR.

In \citet{reiter2019_tadpole}, hereafter Paper~I, we presented optical integral field-unit spectroscopy of the tadpole from the Multi Unit Spectroscopic Explorer (MUSE) on the Very Large Telescope (VLT).  
Physical properties derived from optical diagnostics probe the conditions in the ionized and partially neutral gas on the surface of the globule and in the externally irradiated jet+outflow system.  
Combining these diagnostics with spatially resolved observations of the kinematics in the cold molecular gas provides a powerful probe of how the environment affects the evolution of small globules.

In this paper (Paper~II), we present spatially and spectrally resolved ALMA  observations of the cold, molecular gas in the tadpole.  
Unlike previous observations of the molecular content of small globules, our ALMA data have angular resolution comparable to the \emph{Hubble Space Telescope (HST)}, allowing us to detect the jet-driving source and the associated molecular outflow for the first time. 
Comparing the physical properties of the cold molecular gas derived here with the impact of the environment determined from optical IFU spectroscopy (Paper~I), we will quantify how feedback from the high-mass star-forming environment determines the fate of this small globule (Reiter et al.\ in prep; Paper~III). 

\section{Observations}\label{s:obs}

ALMA Band 7 and 6 observations of the HH~900 globule were obtained in 2016 and 2017, respectively. Table \ref{t:ALMA_obs} lists  observational parameters including the derived flux of the phase calibrator J1047$-$6217, the date of the observations, the on-source time, the maximum angular resolution (MAR, given by $\lambda/L_{\rm max}$, where $L_{\rm max}$ is the largest baseline), and the maximum recoverable scale (MRS, given by $0.6\lambda/L_{\rm min}$, where $L_{\rm min}$ is the shortest baseline, see Eq.\  (3.28) in \citealp{ATH7}). Observations consisted of 12m-array (40 antennae) single pointing scans toward R.A.=$10\fh45\fm19\fs3$, decl.=$-59^\circ44\arcmin23\farcs0$ (ICRS). Band 6 observations were taken using a medium ($L_{max}=1.1$ km, C40-5) and long baseline ($L_{max}=14.9$ km, C40-8) configuration, while Band 7 was observed using only the medium baseline configuration (C40-5). The field-of-view (FOV) FWHM of Band 6 and 7 is 27\arcsec\ and 19,\arcsec respectively. 

Our spectral setup targeted rotational transitions J=2-1 and J=3-2 of the CO isotopologues $^{13}$CO and C$^{18}$O, as well as $^{12}$CO J=2-1 and  SiO J=5-4 in the Band 6 setup. 
We observed $^{13}$CO and C$^{18}$O lines  with velocity  resolution ranging between $0.06$ \kms\ at Band 7 and $0.08$ \kms\ at Band 6. The main CO isotopologue and the SiO line were observed with $0.16$ and $0.34$ \kms\ velocity resolution, respectively. 
We also observed continuum spectral windows with resolution between 1.7--2.5 \kms\ covering approximately 4 and 6 GHz in  Bands 6 and 7, respectively. 
Bandpass, flux, and gain calibration against external calibrators were done using the \emph{Common Astronomy and Software Applications} \citep[CASA,][]{mcm07} v4.7. 
Bandpass and flux calibrators for the Band 6 observations performed in September 2017 were J0635$-$7516 and J0538$-$4405, respectively; while a single source, J1107$-$4449, was used for the May 2017 observations. 
Bandpass and flux calibrators for Band 7 observations were J0538$-$4405 and J1037$-$2934, respectively. Fluxes of the flux calibrators were  interpolated from measurements  
performed by the ALMA calibrator survey separated from our project  by less than two days. The fluxes derived for the bandpass calibrators (when it is not the flux calibrator) 
are within $\sim2\%$ compared with the values given by the ALMA calibrator survey (taken within 3 days from our observations) in all cases.  
Imaging and self-calibration was performed using CASA v5.4. Band 6 images were obtained using \texttt{tclean} by combining the data from both configurations in Table \ref{t:ALMA_obs}. Absolute flux scaling uncertainty is estimated to be about 15\%.

The synthesized beamsizes of the reduced data are given in Table \ref{t:ALMA_lines} and range typically between 0\farcs1--0\farcs2, 
providing an excellent complement to H$\alpha$ images obtained with \emph{HST} \citep[see][]{reiter2015_hh900} and corresponding to a spatial resolution 230--460~AU at the distance of Carina \citep[2.3~kpc;][]{smith2006_distance}.
We describe how we resolve the discrepancy in the ALMA and \emph{HST} astrometry in Appendix~\ref{s:astrometry}.

\begin{table}
\caption{ALMA observation parameters\label{t:ALMA_obs}}
\begin{center}
\begin{tabular}{llllll}
\hline\hline
Band & Pha.\ cal. & Observation & $t_{\rm on}$& MAR$^\star$  & MRS$^\dagger$  \\ 
 & flux & date &  &  &  \\
 & [mJy] & [dd-mm-yyyy] &[s]&[\arcsec] & [\arcsec]\\
\hline\hline
6 & 560.7 & 08-05-2017 & 912 & 0.24 & 10.7\\
  & 564.5 & 25-09-2017 & 3008 & 0.02 & 3.9 \\
7 & 502.2 & 31-10-2016 & 910 & 0.17 & 6.0 \\
\hline
\multicolumn{6}{l}{$^\star$ MAR = maximum angular resolution} \\
\multicolumn{6}{l}{$^\dagger$ MRS = maximum recoverable scale, given by $0.6\lambda/L_{\rm min}$} \\ 
\end{tabular}
\end{center}
\end{table}

\begin{table*}
\caption{Spectral and imaging characteristics of the data.}
\begin{center}
\begin{footnotesize}
\begin{tabular}{lllllllll}
\hline\hline
Name & Frequency & Bandwidth & Resolution & $\theta_{\rm min}$ & $\theta_{\rm max}$ & P.A. & RMS & Comment \\ 
& [GHz] & [MHz] & [km/s] & [\arcsec] & [\arcsec] & [$^\circ$] & [mJy bm$^{-1}$]\\
\hline\hline
\multicolumn{9}{c}{Molecular lines} \\
\hline
SiO J=5-4       & 217.1049800   & 468.75    & 0.337 &  0.099 & 0.101 & -62.2 & 1.08 &\\ 
DCN J=3-2       & 217.2384      & 468.75    & 0.337 &  0.099 & 0.101 & -62.6 & 1.22 & in SiO spectral window  \\ 
\cdo J=2-1 & 219.5603568   & 117.19    & 0.083 &  0.099 & 0.101 & -75.5 & 2.14 & \\ 
\tco J=2-1 & 220.3986765   & 117.19    & 0.083 &  0.097 & 0.126 & -81.3 & 2.69 &\\ 
$^{12}$CO J=2-1 & 230.538       & 234.375   & 0.159 &  0.096 & 0.103 & -77.4 & 2.41 &\\ 
\cdo J=3-2 & 329.3305453   & 117.19    & 0.056 &  0.202 & 0.244 & -26.8 & 12.0 &\\ 
\tco J=3-2 & 330.5879601   & 117.19    & 0.055 &  0.201 & 0.240 & -27.1 & 10.2 &\\ 
CS J=7-6        & 342.8828503   & 1875.0    & 1.71  &  0.209 & 0.225 & -35.4 & 2.69 &\\
\hline
\multicolumn{9}{c}{Continuum} \\
\hline
B6 LSB          & 217.1         & 468.75    & 0.337 &  0.162 & 0.185 & -58.7 & 0.054$^*$ & SiO and DCN lines \\
B6 USB          & 232.2         & 1875.0    & 2.51  &  0.187 & 0.22 & -55.6 & 0.047$^*$ & \\ 
B7 LSB          & 331.6         & 1875.0    & 1.77  &  0.171 & 0.220 & -25.0 & 0.11$^*$ & \\
B7 USB          & 343.0         & 3750.0    & 1.71  & 0.175 & 0.203 & -26.8 & 0.090$^*$ & 2 SpW of $1875$ MHz\\ 
\hline 
\multicolumn{9}{l}{$^*$ RMS of the aggregated bandwidth image. } \\
\end{tabular} 
\end{footnotesize}
\end{center}
\label{t:ALMA_lines}
\end{table*}


\section{Results and Analysis}\label{s:results}

With the superior sensitivity and angular resolution of ALMA, we resolve the structure and kinematics of the cold molecular gas associated with HH~900 jet+outflow system and the tadpole globule for the first time (see Figure~\ref{fig:alma_data}).
Our ALMA observations have angular resolution comparable to \emph{HST}, allowing the most direct 
comparison with the physical structures seen at shorter wavelengths (e.g., Paper~I).

ALMA provides the first look inside the tadpole. 
We detect the previously unseen HH~900 jet-driving source (the \tyso; see 
Section~\ref{ss:dust}). 
These data reveal the bipolar molecular outflow emerging from the \tyso\ that smoothly joins the irradiated outflow at the edge of the globule (see Figure~\ref{fig:landscape} and \ref{fig:oflow_contours}). 
In addition, the largest angular scale of the ALMA observations include a few features with associated molecular emission located outside the tadpole. 
We detect both continuum and CO emission from the YSO that lies in the western limb of the irradiated outflow, PCYC~838 (see Figure~\ref{fig:landscape} and Appendix~\ref{s:pcyc838}). 
A small globulette, located just south of the tadpole tail, can be seen in silhouette against the background nebulosity  and in emission in $^{12}$CO (see Figure~\ref{fig:alma_data}). 
Molecular line data include the serendipitous detection of DCN J=3-2, and CS J=7-6.  
In the following sections, we 
derive the physical properties of these elements of the tadpole and HH~900 jet+outflow system.

\subsection{Optical Depth}\label{ss:tau}

\begin{figure*}
  \centering
  \includegraphics[height=0.9\textheight]{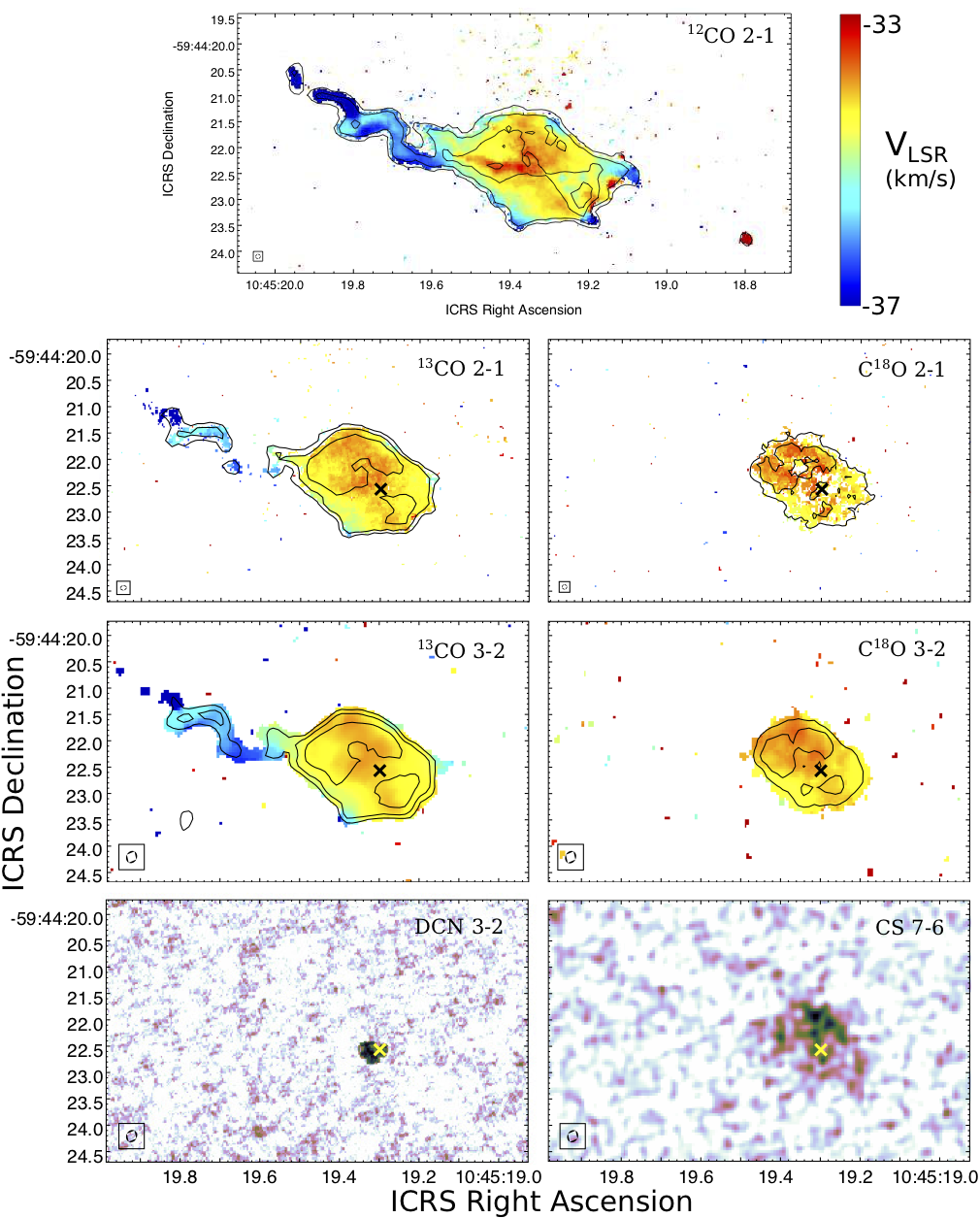}\\
  \caption{
   CO isotopologues panels show moment one maps with moment 0 contours over-plotted. Contours of top panel are the same as in Figure \ref{fig:landscape}. For the other CO isotopologue panels contour levels are $-2$ (dashed), 1, 2, 4, 8, and $16\times\sigma$, with $\sigma$ equal to 5.6 mJy \kms\ and 50 mJy \kms\ for \tco\ J=2-1 and J=3-2; and 5.4 mJy \kms\ and 42 mJy \kms\ for \cdo\ J=2-1 and J=3-2, respectively. Left and right bottom panels use a different color scale and show the moment 0 of DCN J=3-2 and CS J=7-6 lines, respectively.}\label{fig:alma_data} 
\end{figure*}

We have observed multiple isotopologues of CO, allowing us to calculate the optical depth at each position and velocity where emission is significantly detected (see Figure~\ref{fig:alma_data}) as follows: 
\begin{equation}
\frac{T_{\rm main,v}}{T_{\rm iso,v}} = \frac{1 - e^{-\tau_{\rm main,v}}}{1-e^{- \tau_{\rm iso,v}}} = \frac{1-e^{-\tau_{\rm main,v}}}{1-e^{- \tau_{\rm main,v}/R}}
\end{equation}
where we denote the more abundant species as ``main'' and the optically thin transition used to correct it as ``iso'' and assume that the excitation temperature is the same for both molecules. 
To compare the brightness temperature of the CO isotopologues, we convolve each map to the same resolution.
The scale factor $R$ is the relative abundance of the two species; we assume [CO/\tco]$=68\pm20$ and [CO/\cdo]$=570\pm130$ at the  Galactocentric radius of Carina of 8.1 kpc \citep{wilson1999}. 
The brightness temperature ratios $T_{12}/T_{13}$ and $T_{12}/T_{18}$ 
(corresponding to $^{12}$CO/$^{13}$CO and $^{12}$CO/C$^{18}$O emission ratios, respectively)
are both low ($\lesssim 4$) throughout the globule, indicating that $^{12}$CO is very optically thick.

Given the high optical depth in the $^{12}$CO line, we also compute the optical depth of $^{13}$CO by comparing the brightness temperature of $^{13}$CO and C$^{18}$O. 
We find that $^{13}$CO is optically thick with $\tau_{13} >3$ (at the $v_{LSR}$) everywhere that the line is significantly detected.
The fifth column in Table~\ref{t:molecular_props} shows the median optical depth at the source velocity ($-33.5$ \kms) of the optically thick lines. This median is taken from within an ellipse of size $1.2" \times 0.9"$ (P.A.\ $60^{\circ}$) 
centered on the tadpole, roughly corresponding to the lowest contour in the C$^{18}$O J=3-2 panel of Figure~\ref{fig:alma_data}.
We show maps of the spatially-resolved optical depth at the $v_{LSR}$ in Appendix~\ref{s:tau_maps}.

\subsection{Molecular column density}\label{ss:alma_NH}

We compute the column density of each observed transition from the following equation: 
\begin{equation}
  N_{tot} = \frac {4\pi Q(T_{ex}) e^{E_{u} / kT_{ex}} B_{\nu}(T_{ex}) }
    {h c g_u A_{ul} [J_{\nu}(T_{ex}) - J_{\nu}(T_{\rm cmb})] } 
    \int 
    T_{\rm mb} \frac{ \tau_{v}}{1 - e^{-\tau_{v}}} dv  \,\,\mathrm{cm}^{-2}
\end{equation}\label{eq:Ncol}
where
$Q(T_{ex})$ is the rotational partition function for a given excitation temperature 
(the statistical sum over all rotational energy levels),  
$g_u$ is the rotational degeneracy of the upper level with energy $E_u$, 
$A_{ul}$ is the Einstein A coefficient for the transition, 
$k$ is the Boltzmann constant, 
$h$ is the Planck constant, 
$B_{\nu}$ is the Planck function,
$J_{\nu} =(h\nu/k)/[\exp(h\nu/kT)-1]$ is the Planck function in temperature units (K), and 
$\tau_{v} / (1 - e^{-\tau_{v}})$ is a correction factor for non-zero optical depth \citep[see, e.g.,][]{goldsmith1999}. 
We obtain the relevant parameters for each molecule (frequency, rotational partition function, Einstein A coefficients, etc.) from the JPL Spectral Line Catalog \citep{pickett1998} and the Leiden Atomic and Molecular Database \citep[LAMBDA;][]{schoier2005}. 
We report median column densities (using the same extraction region as for the optical depth) in Table~\ref{t:molecular_props}. 
Maps of the spatially-resolved column density calculation for each of the CO isotopologues are shown in Appendix~\ref{s:Ncol_maps}. 

\begin{table*}
\caption{Summary of molecular line derived physical properties. Columns are the species/transition, peak and median intensities, median column density if optically thin, median optical depth, and median column if optically thick, respectively. \label{t:molecular_props}}
\centering
\begin{footnotesize}
\begin{tabular}{lllllll}
\hline\hline
Element & $I_{\mathrm{peak}}$ &
$I_{\mathrm{median}}$ & 
 log(N$_{\mathrm{thin}}$)$_{\mathrm{median}}$ &  $\tau^*_{\mathrm{median}}$ &  log(N$_{\mathrm{thick}}$)$_{\mathrm{peak}}$ & 
 log(N$_{\mathrm{thick}}$)$_{\mathrm{median}}$\\ 
        & [K~km~s$^{-1}$] & [K~km~s$^{-1}$] & 
         [cm$^{-2}$] & & [cm$^{-2}$] & [cm$^{-2}$] \\
\hline
$^{12}$CO J=2-1 & 158.6 & 69.3 & 16.5 & 209 & 20.7 & 19.0 \\ 
\tco\ J=2-1 & 75.1 & 33.5 & 16.2 & 6.0 & 18.4 & 17.1 \\
\cdo\ J=2-1 & 61.7 & 24.0 & 16.1 & \ldots & \ldots & \ldots\\
DCN J=3-2 & 16.9 & 3.1 & 13.1 & \ldots & \ldots & \ldots\\
\tco\ J=3-2 & 67.5 & 40.6 & 16.3 & 4.7 & 18.1 & 17.1 \\
\cdo\ J=3-2 & 31.6  & 15.5  & 15.9  & \ldots & \ldots & \ldots   \\
CS J=7-6    & 13.7 & 7.3 & 13.6 & \ldots & \ldots & \ldots \\
\hline
\multicolumn{7}{l}{$^*$ median value taken at the source velocity, $v_{LSR} = -33.5$~\kms. }\\

\end{tabular}
\end{footnotesize}
\end{table*}
%

To compute the column density, we assume an excitation temperature, $T_{ex}=20$~K.
Assuming a single excitation temperature is a large source of uncertainty as this is often a poor assumption \citep[see discussion in][]{mangum2015}. 
The tadpole is embedded in the brightest portion of the H~{\sc ii} region where temperatures are estimated to be somewhat higher \citep[$\gtrsim 30$~K, see, e.g.,][]{roccatagliata2013}.
At the same time, our data suggest that gas in the center of the tadpole remains cold (see Sections~\ref{s:tadpole_temp} and \ref{ss:dust}). 
Adopting a higher excitation temperature ($T_{ex}\approx40-80$~K) 
or a variable excitation temperature, as described in Section~\ref{s:tadpole_temp}, 
changes our results by a factor of $\lesssim 2$ 
(see Appendix~\ref{s:Ncol_maps} and Figure~\ref{fig:CO_Ncol_maps}).

\subsection{Molecular gas mass}\label{ss:alma_masses}

We estimate the molecular mass of the tadpole from the C$^{18}$O as this is the least optically thick of the CO isotopologues. We compute the mass as follows: 
\begin{equation}
M_{gas} = N({\rm C^{18}O}) \left[ \frac{\rm H_2}{\rm C^{18}O} \right] \mu_g m({\rm H_2}) \pi a_{min} a_{max}
\end{equation}
where
$\left[\mathrm{H_2}/\mathrm{C^{18}O} \right] = 4.85 \times 10^6$ is the abundance of H$_2$ compared to C$^{18}$O, 
$\mu_g = 1.36$ is the mean molecular weight, 
$m(H_2)$ is the mass of molecular hydrogen, and 
$a_{min} \approx 0.9^{\prime\prime}$ and $a_{max} \approx 1.2^{\prime\prime}$ are the minor and major axes of the tadpole, respectively. 
This gives a gas mass of $\sim 0.6$~M$_{\odot}$.

We use the C$^{18}$O column density to estimate the mass since both $^{12}$CO and $^{13}$CO are optically thick. 
However, in regions bathed in ionizing radiation, isotope-selective photodissociation may alter the relative abundance of optically thin isotopologues like C$^{18}$O \citep[e.g.,][]{keene1998} 
If isotope-selective photodissociation has reduced the abundance of C$^{18}$O, then this mass estimate will be a lower limit. This will also be the case if C$^{18}$O is optically thick. 

We also estimate the mass of the small globulette located just below the tadpole tail (see Figures~\ref{fig:landscape} and \ref{fig:oflow_contours}). 
We compute the globulette mass using the optically thin $^{12}$CO emission (the globulette is not detected in C$^{18}$O). 
Using the abundance of H$_2$ compared to $^{12}$CO, 
$\left[ \mathrm{H_2} / \mathrm{^{12}CO} \right] = 10^4$ and again assuming $T_{ex}=20$~K, we estimate a mass M$_{\mathrm{globulette}} \approx 0.04$~M$_{\mathrm{Jupiter}}$. 
This is a factor of a few smaller than the typical globulette mass in the \citet{grenman2014} catalog (this object is not in that sample).

\subsection{CO emission toward the externally heated tadpole
}\label{s:tadpole_temp}
\begin{figure}
  \centering
    \includegraphics[width=0.475\textwidth]{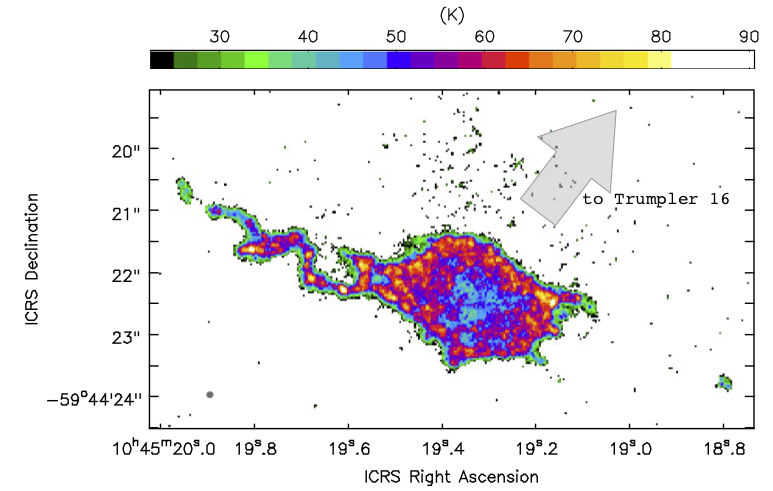}%
    \caption{Peak brightness temperature of $^{12}$CO J=2-1.\label{fig:bTemp}}
\end{figure}

We detect CO emission throughout the tadpole and measure high optical depths (see Table~\ref{t:molecular_props}) that suggest that none of the isotopologues probe gas in the immediate environs of the \tyso\ (see Figure~\ref{fig:alma_data}). 
For the tadpole, 
the different appearance between the continuum (see Section~\ref{ss:dust}) and the CO integrated lines (compare Figures~\ref{fig:alma_data} and \ref{fig:cont}) can be explained mostly by differences in optical depth.  
Assuming [H$_2$/CO]$=10^{4}$ and a linewidth of 1 \kms, the ratio between the peak optical depth of the CO J=2-1 line 
and that of dust is approximately $10^{5}$. This means that a tadpole mass $\gtrsim 0.01~M_{\sun}$ is sufficient to ensure optically thick CO conditions. Therefore, we can use the Eddington-Barbier approximation and assume that Figure \ref{fig:bTemp} is showing the temperature of the tadpole  at different depths. Because lines of sight directed farther from the center of the tadpole probe layers 
located at larger radii, the limb-brightening shown in Figure \ref{fig:bTemp} is  interpreted as a positive radial ($\frac{dT}{dr}>0$) temperature gradient (external heating). This gradient seems to characterize the tadpole at least as deep as the CO line can probe.

Limb brightening is the main evidence for a positive radial temperature gradient. 
In this spatially-resolved source, lines of sight closest to the globule edge trace primarily surface material where we measure a larger brightness temperature. 
Two additional lines of evidence support this interpretation.

Less abundant isotopologues like \tco\ and \cdo\ can trace deeper, and therefore, probe colder material in the tadpole than $^{12}$CO. 
Assuming optically thick emission, 
this means that the $\tau=1$ surface will be deeper in the globule for rarer isotopologues. 
If this is the case, the observation that 
$T_b(^{12}CO) > T_b (^{13}CO) > T_b(C^{18}O)$  
 (see Appendix~\ref{s:Tmaps} and the peak brightness temperature maps in Figure~\ref{fig:Tmaps}) indicates a thermal gradient. 
We compute high optical depths for both $^{12}$CO and \tco\ (see Section~\ref{ss:tau}). 
We do not have the data to constrain the optical depth of the \cdo, although it may be optically thin along lines of sight away from the \tyso. 
We note that higher brightness temperatures from more abundant species is also expected under optically thin conditions.

\begin{figure}
  \centering%
    $\begin{array}{c}
    \includegraphics[trim=0mm 0mm 0mm 0mm,angle=0,scale=0.45]{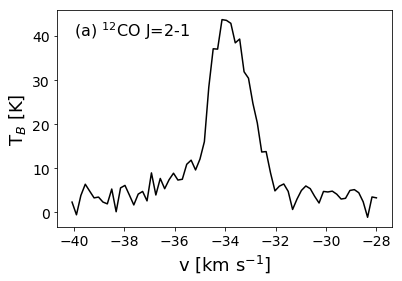} \\
    \includegraphics[trim=0mm 0mm 0mm 0mm,angle=0,scale=0.45]{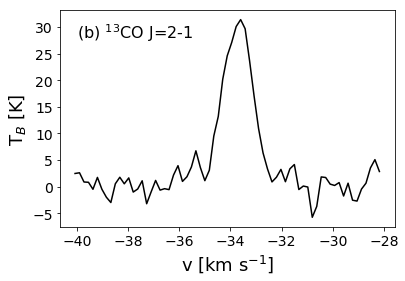} \\
    \includegraphics[trim=0mm 0mm 0mm 0mm,angle=0,scale=0.45]{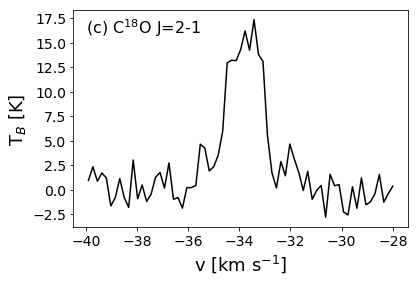} \\
    \includegraphics[trim=0mm 0mm 0mm 0mm,angle=0,scale=0.45]{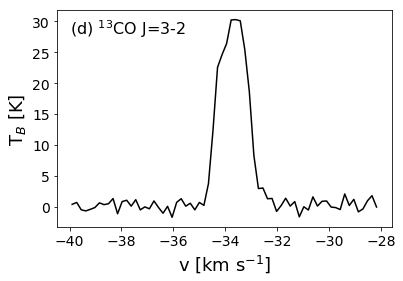} \\
    \includegraphics[trim=0mm 0mm 0mm 0mm,angle=0,scale=0.45]{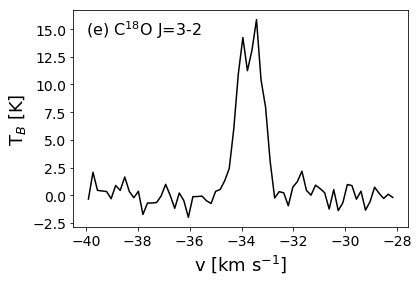} \\
    \end{array}$
\caption{Median CO line profiles near the \tyso. Panels (a) to (e) display the molecule and transition in their top left corner. 
  \label{fig:modelCO}}
\end{figure}

The J=2-1 to J=3-2 line ratios also suggest that all three isotopologues are optically thick. 
We show the median intensity of each CO isotopologue toward the \tyso\ in Figure~\ref{fig:modelCO}. 
The median is taken within a circle of radius $0\farcs3$ centred on the position listed in Table \ref{t:fluxes_continuum}. 
The peak temperature of \tco\ and \cdo\ are similar for both the J=2-1 and J=3-2 transitions. 
This is expected for optically thick lines as $T_{mb} \approx T_{ex}$ when $\tau >>1$ (the opacities for both transitions are within a factor of two for 8~K$<T_{ex}<$100~K). 
Optically thin emission at $T_{ex} \sim 16$~K would also predict similar peaks for the J=2-1 and J=3-2 lines in local thermodynamic equilibrium (LTE). 
However, $T_b(^{13}CO)> 16$~K, indicating optically thick emission.

One additional notable feature of the CO lines is that, despite their high optical depth, their shapes do not show the self-absorbed profiles usually observed in CO toward star formation regions. 
Indeed, toward the tadpole, the  profiles shown in Figure \ref{fig:modelCO} display a single peak and are roughly symmetric, reminiscent of optically thin lines.

Line emission from a cloud with a temperature gradient is characterized by different parts of the line tracing material with different excitation temperatures. At the line peak, where line opacity is highest, the emission is tracing on average more external layers of the core compared with emission at velocities in the sloping wings of the line.
The specifics of the thermal gradient depend on the incident radiation and the density profile of the cloud. 
Nevertheless, a positive radial thermal profile can explain the lack of self-absorption: intervening material in outer layers is warmer, so it does not decrease the intensity of lines arising from colder inner regions. 

\subsection{The molecular outflow associated with the irradiated HH~900 jet+outflow system}\label{ss:jet_props}
\begin{figure*}
   \hspace*{-2em} 
\includegraphics[width=0.55\textwidth]{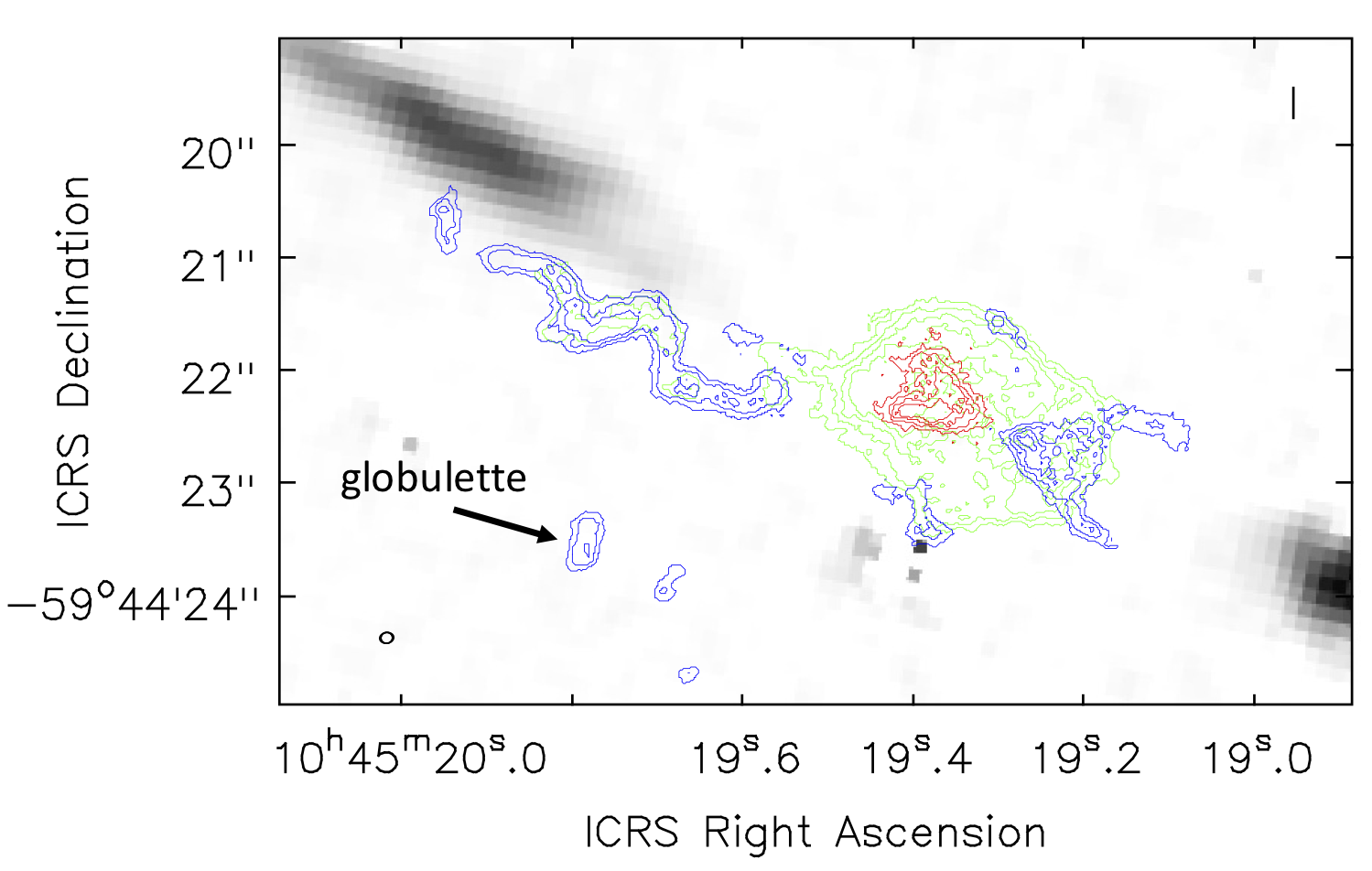}%
    \includegraphics[width=0.5\textwidth]{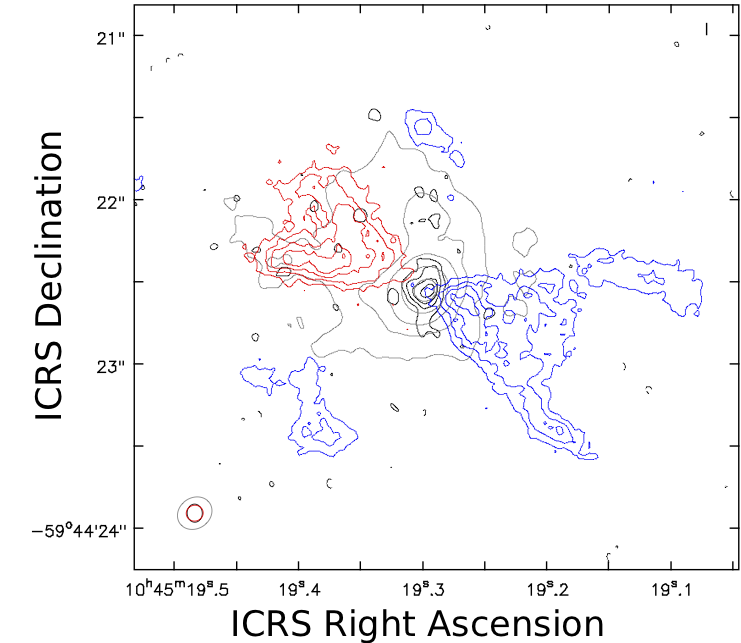}
  \caption{Red and blue contours in both panels show the integrated emission from the wings of the CO J=2-1 line (using $[-57,-37]$ \kms\ and  $[-30,-11]$ \kms, for the blue and red lobes, respectively). Beams are shown in the lower-left corner of each panel. Dashed contours mark negative levels. Contour levels: blue: $-3$, 7, 13, 19, and $25\times\sigma=3.7$ K \kms; red: $-3$, 8, 16, 24, and $32\times\sigma=3.3$ K \kms.
    \textit{Left:} Gray-scale: [Fe~{\sc ii}] emission from HST as shown in \citep{reiter2015_hh900}. Green contours show the integrated  \tco\ J=2-1 line. Levels: $-3$, 15, 30, 45, $60\times\sigma=1$ K \kms.
    \textit{Right:} Zoom in into the outflow region. Black and gray contours show the continuum emission at 232.2 GHz imaged using robust  weighting parameters $-0.5$ and $1.0$, respectively.  Black levels: $-3$, 3, 5, 7, and $10\times\sigma=108$ $\mu$Jy beam$^{-1}$; gray levels: $-5$, 5, 10, 15, 25, and $40\times\sigma=52$ $\mu$Jy beam$^{-1}$.
\label{fig:oflow_contours}} 
\end{figure*}

\begin{figure}
  \includegraphics[width=0.5\textwidth]{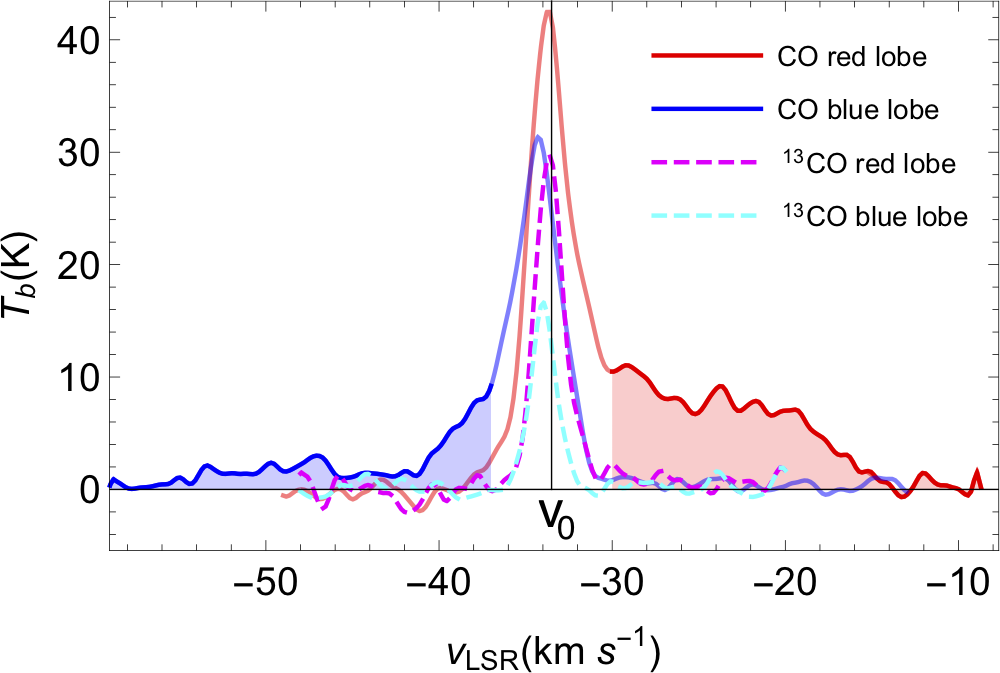}
  \caption{Profiles of the average CO and \tco\ J=2-1 emission within  the red- and blue-shifted lobes of the biconical outflow detected toward the \tyso\  YSO. A vertical line marks the assumed central $v_{\rm LSR}$ of $-33.5~\kms$.  Shaded regions indicate the blue- and red-shifted wing emission extending between $[-57,-37]$ \kms\ and  $[-30,-11]$ \kms, respectively. 
  }\label{fig:oflow_profile}
\end{figure}

\begin{figure*}
  \centering
  $\begin{array}{cc}
    \includegraphics[trim=10mm 7mm 0mm 20mm,angle=0,scale=0.345]{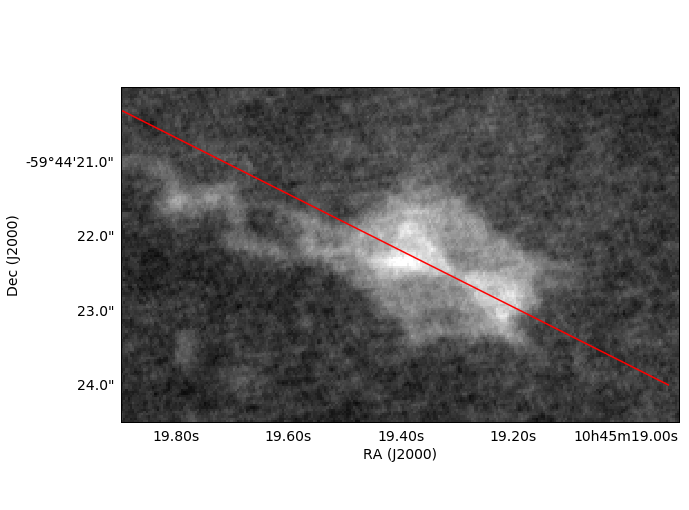} &
    \includegraphics[trim=0mm 35mm 10mm 20mm,angle=0,scale=0.525]{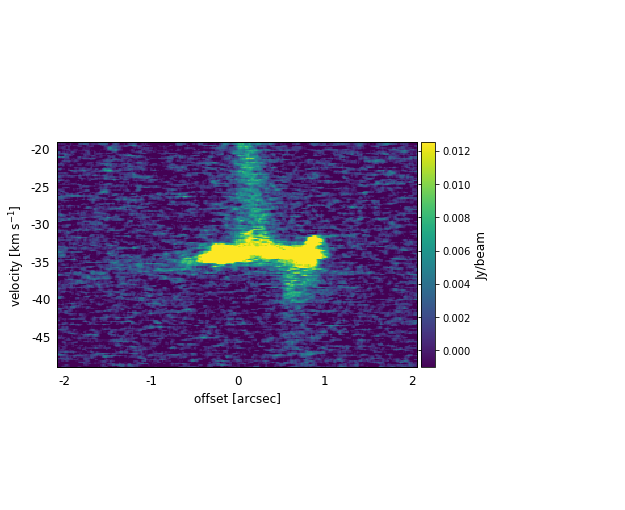} \\
    \end{array}$
  \caption{Tracing location (\textit{left}) used to make the position-velocity diagram (\textit{right}) of the $^{12}$CO J=2-1 showing the molecular outflow associated with HH~900. Emission is summed in an $0\farcs25$ aperture. 
}\label{fig:pv_diagram} 
\end{figure*}

We report the first detection of the cold, molecular outflow associated with the HH~900 jet+outflow system (see Figures~\ref{fig:oflow_contours} and \ref{fig:oflow_profile}).
Two biconical outflow cavities (opening angles of $\sim 50^{\circ}$) emerge from the protostar detected in the globule (discussed in the next Section).
The lobes are redshifted and blueshifted in the same sense as the irradiated jet+outflow components seen outside the globule 
(see Figure~\ref{fig:pv_diagram}). 
The biconical cavities open to the same width as the globule at the edge.
Structure in the blueshifted $^{12}$CO J=2-1 emission traces the uneven edge of the globule, protruding beyond the boundaries of the globule delineated by the optically thin isotopologues (see Figure~\ref{fig:oflow_contours}). 
These extensions from the outflow cone 
overlap with two of the three small dust streamers seen in silhouette in the \emph{HST} images (see Figure~\ref{fig:landscape}). 
\citet{smith2010} suggested that these streamers are 
limb-darkened by dust in the side walls of the outflow cavity. 
A third streamer, located closer to the major axis of the globule (see Figure~\ref{fig:landscape}), is also seen in CO but does not appear to be blueshifted (see Figure~\ref{fig:oflow_contours}).  
In this environment, the molecular outflow is only detected within the protection of the high-density globule.
The outflow is not seen in either transition of $^{13}$CO or C$^{18}$O.

To estimate the physical parameters of the outflowing gas,
we integrate the line emission in regions enclosed within the lowest contour in Figure \ref{fig:oflow_contours} 
(as described in the figure caption, these are 26.4~K~\kms\ for the red lobe and 25.9~K~\kms\ for the blue lobe). 
In the blue-shifted case, we integrate only inside the roughly conical/triangular shape extending to the southwest from the continuum source. 
Figure \ref{fig:oflow_profile} shows the average $^{12}$CO and \tco\ J=2-1 emission from these two regions. Red-shifted wing emission is prominent, extending up to $+19$~\kms\ from the globule's $v_{\rm LSR}$ ($v_0=-33.5~\kms$). The blue-shifted side displays evident
wing emission starting from  $(v_0-3.5)$ \kms\ up to radial velocities of $(v_0-24)$ \kms.

To evaluate mass, momentum, and energy contained in the outflow lobes we use the velocity moments of the CO J=2-1 line following the methods described in \citet{calvet1983}. We directly integrate the profile in the line wings ---  $|v-v_0|>3.5~\kms$ --- and correct for the low velocity outflow material following  \citet{margulis1985}.
We assume optically thin conditions for the high-velocity wing emission, a single excitation temperature $T_{\rm ex}=100$ K, and a [H$_2$/CO] abundance ratio of $10^{4}$. The adopted excitation temperature is consistent with that found toward outflow gas in low- and intermediate-mass stars \citep{van09,guz11,yil12,gom19}.
As shown in Figure \ref{fig:oflow_profile}, \tco\ is not detected in the wings, consistent with CO opacities $\le0.1$ in the outflow. 
The obtained physical parameters, uncorrected for inclination, are given in Table~\ref{t:outflow_props}.

\citet{reiter2015_hh900} estimated that the jet is tilted $\lesssim 10^{\circ}$ from the plane of the sky (corresponding to an inclination angle $i \gtrsim 80^{\circ}$, where $i=90^{\circ}$ lies in the plane of the sky). 
Assuming that all outflowing material is directed along the outflow axis, the inclination corrections for the momentum, energy, 
mass-loss rate, 
and momentum rate are $(\cos i)^{-1}$, $(\cos i)^{-2}$, 
$\tan i$, 
and $(\cos i)^{-2}\sin i$, respectively, where $i$ is the inclination angle between the outflow axis and the line of sight. However, we caution that a large fraction of the material in molecular outflows moves in directions transverse to the outflow axis, thus making these naive inclination corrections leads to an overestimate of the outflow parameters when compared to simulations \citep{downes2007}. Furthermore, considering that the semi-opening angle of the outflow is $27.5^\circ$ and that there is no discernible blue- or red-shifted emission associated with the opposite lobe, following \citet{cabrit1986} we conclude that $i\le62.5^\circ$. This is somewhat less than the $i \gtrsim 80^{\circ}$ estimated in \citet{reiter2015_hh900}. 
Uncertainties in the velocities, and therefore inclination estimates, determined from optical spectroscopy are large as they are measured much less precisely than those from millimeter emission lines. 
Accounting for an uncertainty of $\sim 5^{\circ}$ (from the optical), the inclination estimates remain discrepant, leaving open the possibility that the jet and the molecular outflow have different inclination angles. 
Using $T_{\rm ex}=50$~K decreases the mass estimate by $\sim40\%$. The uncertainties  include those derived from the noise of the data and the flux scaling, but they do not reflect the systematic uncertainties such as distance, inclination, abundance variations, or fraction of outflow material confused within the local $v_{\rm LSR}$.

Our spectral setup included SiO J=5-4 which has been observed to trace the collimated, high-velocity jet in some cases \citep[e.g.,][]{codella2007,leurini2013,codella2013}.
However, this line is not detected anywhere in our ALMA map. 
We note that the upper energy level of SiO J=5-4 is a factor of $\sim 2$ higher than $^{12}$CO, leaving open the possibility that lower J transitions of SiO may be detectable. 
However, the lack of SiO emission is consistent with the relatively low outflow velocity. 
Gas phase SiO is thought to be produced when Si is removed from grains by either sputtering or grain-grain collisions. 
To produce the observed column densities of SiO in molecular outflow regions requires shock velocities $\approx25$ \kms\ and densities of the order 10$^5$~cm$^{-3}$ for sputtering \citep{schilke1997,gusdorf08} or grain-grain collisions \citep{caselli1997}.

Combining the outflow length, $R_{lobe}$, and velocity yields the dynamical age, $t_{dyn} = R_{lobe} / v$. 
We use the dynamical age to calculate the outflow mass-loss rate, $\dot{M}_{out} = M_{out} / t_{dyn}$. 
To estimate the momentum of a jet-driven molecular outflow, we follow the prescriptions of \citet{downes2007} and use the quantities in Table \ref{t:outflow_props} without any inclination correction. Note also that from Figure~\ref{fig:oflow_profile} both red-shifted outflowing material in the blue-shifted outflow lobe and vice versa are either negligible or 
they appear entirely confused with the material at $v_0$. For the  force we follow their ``perpendicular'' method with 
$R_{\rm lobe}=0\farcs6\times d=1380$~AU for both lobes.
Using the velocities in Table \ref{t:outflow_props}, we get
\begin{align*}
    &\dot{P}_{\rm blue}=\frac{1}{3}\frac{P_{\rm blue}}{|R_{\rm lobe}/(v_{\rm blue}\!-v_0)|}=-6.5\times10^{-5}~\text{M}_{\sun}\text{km s}^{-1}~\text{yr}^{-1}~~,\\
    &\dot{P}_{\rm red}=\frac{1}{3}\frac{P_{\rm red}}{|R_{\rm lobe}/(v_{\rm red}-v_0)|}=8.5\times10^{-5}~\text{M}_{\sun}\text{km s}^{-1}~\text{yr}^{-1}~~.
\end{align*}


\begin{table*}
\caption{Summary of the physical properties of the outflow. 
For the red and blue lobes of the outflow, columns are the mass of outflow, velocity, momentum, kinetic energy, length of the lobe, dynamical time, mass-loss rate, and momentum flux. 
\label{t:outflow_props}}
\centering
\begin{footnotesize}
\begin{tabular}{lllllllll}
\hline\hline
Lobe    & $M_{out}$                   & $v-v_0$       & $P$                               & K.E.              & R$_{lobe}$ &t$_{\mathrm{dyn}}$ & $\dot{M}_{out}$ & $\dot{P}$ \\ 
        &[$10^{-3}$ \Msun ]&[\kms]  & [$10^{-2}\Msun\kms$] & [$10^{40}$ erg]   &[pc]& [yr]& [\Msun~yr$^{-1}$] & [\Msun~yr$^{-1}$~\kms] \\
\hline
blue & $3.36\pm0.5$ & $-6.45$ & $-2.17\pm0.3$ & $260\pm70$ & 0.007 & 1060 & $3.2\times10^{-6}$ & $-6.5 \times 10^{-5}$ \\ 
red  & $3.15\pm0.5$ & $7.96$  & $2.51\pm0.3$  & $280\pm70$ & 0.007 & 860 & $3.7\times10^{-6}$ & $8.5 \times 10^{-5}$ \\
\hline
\multicolumn{9}{l}{Using $T_{\rm ex}=100$  K. Quantities are not corrected for inclination or opacity.} \\
\end{tabular}
\end{footnotesize}
\end{table*}

%

\subsection{Dust continuum emission}\label{ss:dust}

\begin{figure*}
\centering%
\includegraphics[width=\textwidth]{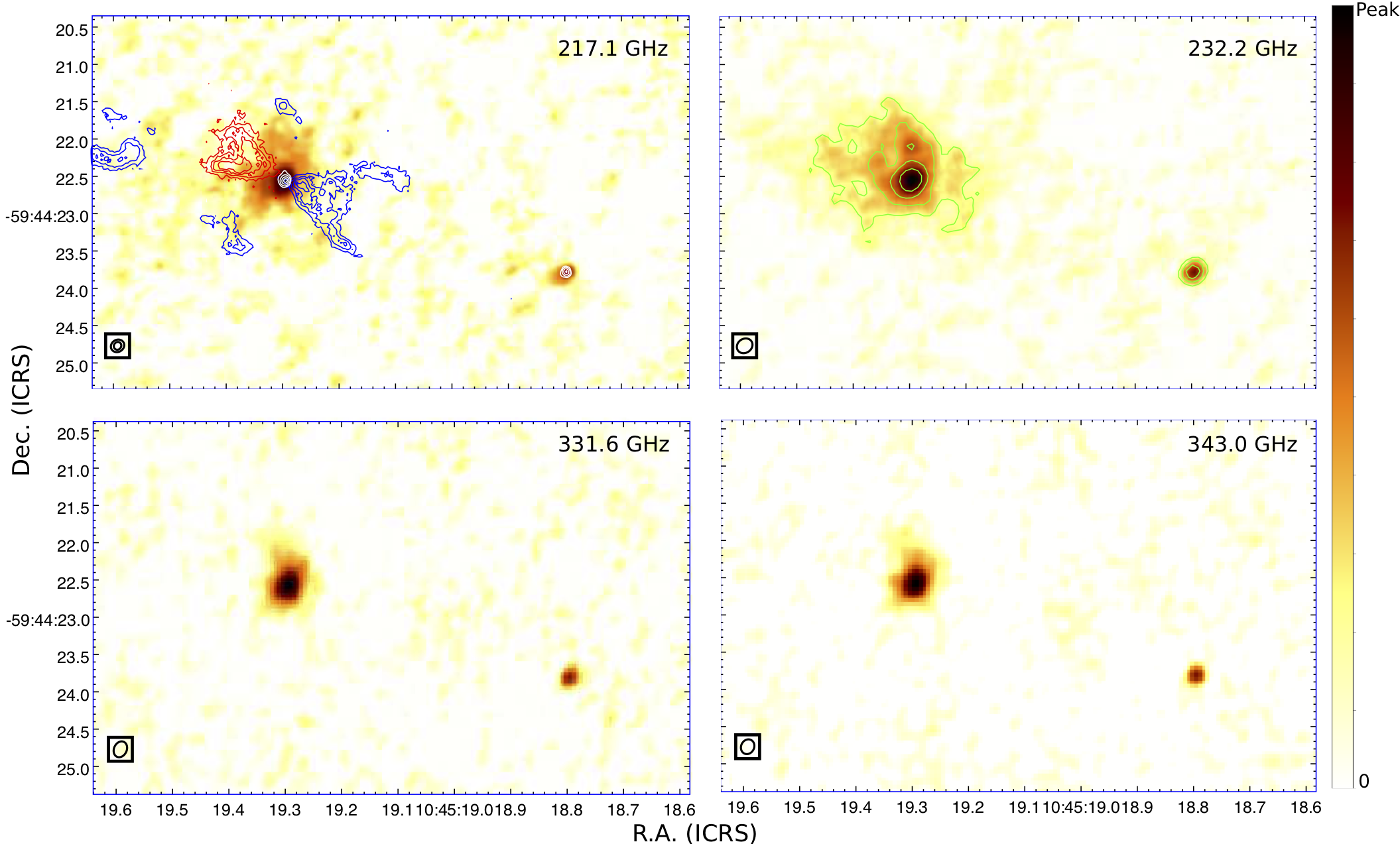} 
\caption{Continuum emission at the four frequencies listed in Table \ref{t:fluxes_continuum}. Beams are indicated in the bottom-left corner of each panel. \emph{Top left panel.} Peak: 1.58 mJy. White contours at 4, 6, 9, 12, and $15\times\sigma=0.82$ mJy beam$^{-1}$ show the 232.2 GHz continuum imaged using only long ($>400~{\rm k}\lambda$) baselines. We also display the same red and blue contour levels from CO J=2-1 wing emission as in Figure \ref{fig:oflow_contours}. \emph{Top right panel.} Peak: 2.43 mJy beam$^{-1}$. The five green contours levels are linearly spaced between $4\times\sigma=0.47$ mJy beam$^{-1}$ and the peak. 
\emph{Bottom left panel.} Peak: 5.32 mJy beam$^{-1}$. \emph{Bottom right panel.} Peak: 5.68 mJy beam$^{-1}$.
}\label{fig:cont} 
\end{figure*}

We detect two point-like sources in each of the continuum bands observed. 
Figure~\ref{fig:cont} shows the two point sources; 
the first resides inside the tadpole globule, on the jet axis and at the origin of the molecular outflow (see Section~\ref{ss:jet_props}).
We propose that this source is 
the young stellar object (YSO) that drives the HH~900 jet+outflow system (the \tyso), seen for the first time with ALMA (Figure~\ref{fig:oflow_contours}).
The second continuum source coincides with the YSO that lies in the western limb of the HH~900 outflow. 
\citet{povich2011} identified this object as a candidate YSO, PCYC~838, based on model fits to the IR SED (see Figure~\ref{fig:landscape} and Appendix~\ref{s:pcyc838}). 
A third star, just beneath the southern edge of the tadpole globule, is not detected with ALMA (PCYC~842, see Figure~\ref{fig:landscape}).

\begin{table*}
\caption{Position and flux densities from the compact continuum source.\label{t:fluxes_continuum}}
\centering
\begin{footnotesize}
\begin{tabular}{llllllll}
\hline\hline
Source & R.A. & decl.&  217.1~GHz & 232.2~GHz &  331.6~GHz & 343.0~GHz & $\alpha^{\dagger}$\\
        & (J2000) & (J2000) & [mJy] & [mJy] & [mJy] & [mJy] & \\
        \hline
\tyso$^a$  &10:45:19.296 & $-$59:44:22.55 & 4.70 & 5.86 & 12.54 & 13.48 & $2.2\pm0.5$\\ 
tadpole head  & ... & ... & 25.2 & 38.7 & 16.8 & 14.9 & ... \\ 
\hline
\multicolumn{8}{l}{$^{\dagger}$Spectral index of best power-law fit to flux densities ($S_\nu\propto\nu^\alpha$).}\\
\multicolumn{8}{l}{$^a$Within a radius of 0\farcs3 from peak.}
\end{tabular}
\end{footnotesize}
\end{table*}

%

Figure~\ref{fig:cont} shows that most of the continuum emission from the tadpole comes from the compact source whose central position 
is given in Table \ref{t:fluxes_continuum}. 
This source, the  \tyso, is slightly offset ($0\farcs5$) from the center of the tadpole's head, which  forms an extended envelope around it. 
The tadpole's head is most evident in the CO lines and visible as faint extended emission in the Band 6 images (with highest sensitivity). 

Table \ref{t:fluxes_continuum} gives the flux densities arising from a region within a radius of 0\farcs3 (corresponding to $r_0=690$~AU at the distance to Carina) from the \tyso\ 
as well as the spectral index  of a power-law fit to these fluxes.
We obtain the best-fit model  by minimizing the squared differences between the model and the fluxes, weighted by the inverse of their squared uncertainties. 
Assuming  flux densities  have an  uncertainty of 15\%, we derive   1$\sigma$ error bars in the spectral index  of
$\pm0.5$ \citep[following, e.g.,][]{lampton1976}. These error bars are due in part to the frequency span of our observations, with adjacent bands (6 and 7) not being optimal to estimate accurate spectral indices.

Nevertheless, the derived spectral indices are lower than those expected from optically thin thermal dust emission in the Rayleigh-Jeans limit --- given by $\beta+2$ ---  
assuming that the dust absorption coefficient behaves as $\kappa_\nu=\kappa_0(\nu/\nu_0)^\beta$. Because the 
brightness temperature of our highest resolution continuum images peak at about $\sim4\textrm{-}5$ K, 
it is not likely that the fluxes are dominated by optically thick dust emission. Therefore, the remaining option to explain the flat spectrum must be a combination of low temperatures (making Rayleigh-Jeans less applicable) and low $\beta$. 
Relatively flat mm SEDs are not uncommon in dense regions of star formation \citep[e.g.,][]{orozco2017,guzman2014}, and they are usually attributed to a $\beta$ parameter lower than that of the diffuse ISM dust (characterized by $\beta\ga1.5$) due to dust coagulation \citep{draine2006}.

We model the fluxes toward the \tyso\ as arising from a spherical dusty core with a density profile $\rho\propto r^{-2}$, representing the 
outer envelope of an accreting young star \citep{mckee2007}. 
We assume a dust absorption coefficient as described above  with $\nu_0=230.61$ GHz (1.3 mm), 
$\kappa_0=0.8$ cm$^{2}$~g$^{-1}$ \citep{oss94},  a $\beta=1.0$, and a gas-to-dust mass ratio of 100.
We assume a temperature profile $T(r)=T_0*(r/r_0)^{-0.4}$, where $r_0=690$~AU. 
We expect this radially decreasing temperature gradient near the \tyso\ (this model is for emission within a radius of 690~AU ($0\farcs3$) of the YSO), although this gradient  reverses on larger scales due to the external influence of environment (see Section~\ref{s:tadpole_temp} and Paper~I).
This profile is  
characteristic of an optically thin, centrally illuminated dusty core with $\beta=1.0$ \citep{adams1985}.

\begin{figure}
  \centering%
  \includegraphics[width=0.5\textwidth]{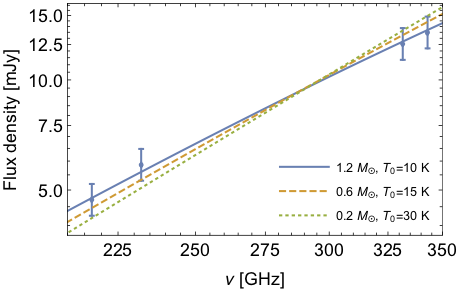}
  \caption{Spectral energy distribution and dust emission models adjusted to the fluxes measured toward \tyso\ within a radius of 0\farcs3 from the peak continuum position (Table \ref{t:fluxes_continuum}).\label{fig:central_SED}}
\end{figure}

Figure~\ref{fig:central_SED} shows best-fit models and derived masses of the core assuming different $T_0$ values.
For comparison,  using an homogeneous temperature of 15 K and the formula \mbox{$M=S_\nu d^2/B_\nu(T)\kappa_\nu$}
we obtain $1.0~M_{\sun}$ for the compact \tyso\ core. 
Given the limited sampling of the continuum spectral energy distribution (SED), we do not attempt to fit a (poorly constrained) blackbody to estimate the luminosity and therefore infer a mass of the \tyso. 
For a young jet-driving source, we expect the SED to peak at wavelengths shorter than we
have observed with ALMA. 
Unfortunately, the \tyso\ is confused with
the two other protostars near the globule in existing data \citep[from $70-500$~\micron\ with e.g., \emph{Herschel}, see][]{ohl12}.

To compute the mass of the entire globule, we use only the Band 6 continuum emission.  
The size of the tadpole head is a significant fraction of the size of the MRS of the Band 7 images (see Table~\ref{t:ALMA_obs}), making it difficult to recover extended, low-surface-brightness emission accurately during deconvolution. All reported fluxes are likely lower bounds (see discussion in Appendix~\ref{s:filtering} and Figure~\ref{fig:compar_ALMA_APEX}). 
We adopt a higher temperature for the extended emission around the compact source (45~K, see Section~\ref{s:tadpole_temp}) to compute a mass of $1.9~M_{\sun}$ for the entire tadpole.
If we instead assume 15~K, as for the compact source, the estimated tadpole mass is $7.3$~\Msun.
We make the conservative choice to adopt a mass of 1.9~\Msun\ for the remainder of our analysis, although this is likely to be a lower bound.

\subsection{Extended emission near the \tyso}\label{ss:circumstellar}
Continuum emission from the HH~900 driving source appears to be marginally resolved at 217~GHz and 232~GHz.
The major axis extends perpendicular to the jet axis.
We estimate an inclination angle $i \lesssim 62.5^{\circ}$  (see Section~\ref{ss:jet_props}) 
for the HH~900 outflow. 
With this orientation, the circumstellar disk around the HH~900 jet-driving source should be viewed nearly edge-on.

Emission lines targeted in this study have been used to measure the size and kinematics of gas disks around low-mass stars \citep[e.g.,][]{andrews2012,tobin2012,tobin2015,ansdell2018}. 
Despite the favorable orientation, the CO lines are too optically thick to probe the kinematic structure in the circumstellar material.
A higher excitation tracer like CH$_3$CN \citep[e.g.,][]{oya2016} may provide more information on the kinematics of the gas closest to the protostar. 

\begin{figure}
  \centering
    \includegraphics[width=0.5\textwidth, trim={4mm 15mm 15mm 10mm},clip]{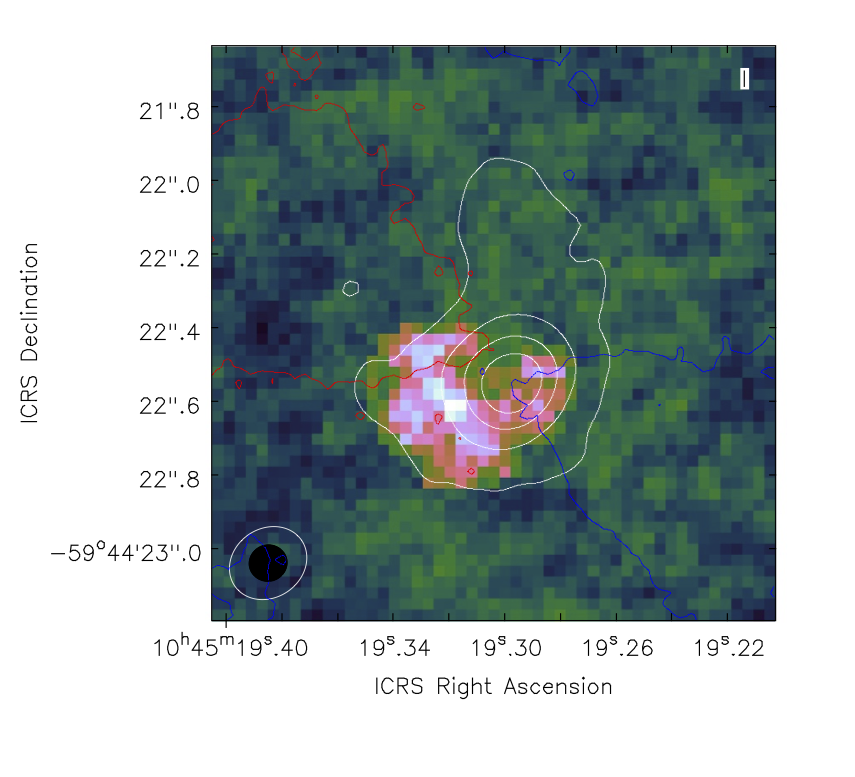}%
  \caption{Integrated DCN emission toward the tadpole. Peak: 16.9 K \kms. White contours show the 232.2 GHz continuum emission. Red and blue outflow contours show the first positive level displayed in Figure \ref{fig:oflow_contours}.
\label{fig:DCN_mom0}} 
\end{figure}
We also report the serendipitous detection of DCN J=3-2 (see Table~\ref{t:molecular_props}). 
Figure \ref{fig:DCN_mom0} shows the integrated 
emission toward the tadpole.
DCN is observed near the location of the protostar,  but forms an asymmetric envelope that seems to wrap the \tyso\ from the south. 
DCN is usually detected in cold gas \citep[e.g.][]{caselli2012}. 
In their study of high-mass star-forming regions, \citet{gerner2015} find a higher fraction of deuterated molecules (including DCN) in colder, less luminous regions. 
At low temperatures ($\lesssim 10$~K) and high densities \citep[$\gtrsim 3 \times 10^4$~cm$^{-3}$, see e.g.,][]{bacmann2002}, CO freezes out of the gas phase, enabling deuterium fractionation \citep{bergin2007}. 
While we do not have a direct measure of the gas temperature near the protostar, we take the detection of DCN as evidence that the gas is cold, creating favorable conditions for the formation of DCN. 
DCN is typically detected toward regions of H$_2$ column density (as estimated from the continuum) of $\gtrsim 3 \times 23$~cm$^{-2}$. 
The typical abundance of HCN ranges between $1 \times 10^{-7}$ and $1 \times 10^{-9}$ \citep{roberts2002}. 
In the absence of fractionation, [DCN/HCN]$\sim 1 \times 10^{-5}$, implying a column density of DCN $<1 \times 10^{12}$~cm$^{-2}$, associated with a line peak $< 1$~K. 
We observe $15~K$ and estimate a DCN column density $\gtrsim 1 \times 10^{13}$~cm$^{-2}$ (see Table~\ref{t:molecular_props}). This is typical: the detection of the deuterated species guarantees some fractionation. 

\subsection{Virial mass estimate}\label{ss:virial} 

We estimate the virial mass of the tadpole globule as a whole assuming that it is a self-gravitating sphere. 
We compute the virial mass using the following expression: 
\begin{equation}
 M_{vir} = \frac{3(5-2n)}{8(3-n)\ln(2)} \frac{(\Delta V)^2R}{G}~~,\label{eq:virial} 
\end{equation}
where 
$n$ is the exponent of the density profile ($\rho \propto r^{-n}$), 
$\Delta V =1.4$~\kms\ is the linewidth of C$^{18}$O J=2-1, 
$R=3000$~AU is the mean radius of the cloud,  
and 
$G$ is the gravitational constant \citep[see e.g.,][]{maclaren1988}. 
Correction factors for non-spherical clouds change Equation~\ref{eq:virial} by less than 10\% \citep{bertoldi1992}.
We consider two density profiles for the globule. 
For a density profile that decreases with radius, 
$\rho \propto r^{-2}$, 
the estimated virial mass is $\sim 3.6$~\Msun. 
For a constant density profile, 
$\rho = constant$, 
this estimate increases to $6.0$~\Msun, 
similar to the mass (7.3~\Msun) we derive assuming cold dust (T=15~K) in Section~\ref{ss:dust}. 
As a gravitational mass, these estimates include the mass of the \tyso. 
Neither estimate includes the pressure of the external environment, although the physical parameters derived in Paper~I suggest that this will be important. 
We leave a more complete exploration of the effect of the external environment to Paper~III.

\subsection{Cold gas kinematics}\label{ss:gas_kin} 
\begin{figure*}
  \centering
$\begin{array}{c}
    \includegraphics[trim=70mm 40mm 0mm 30mm,angle=0,scale=0.475]{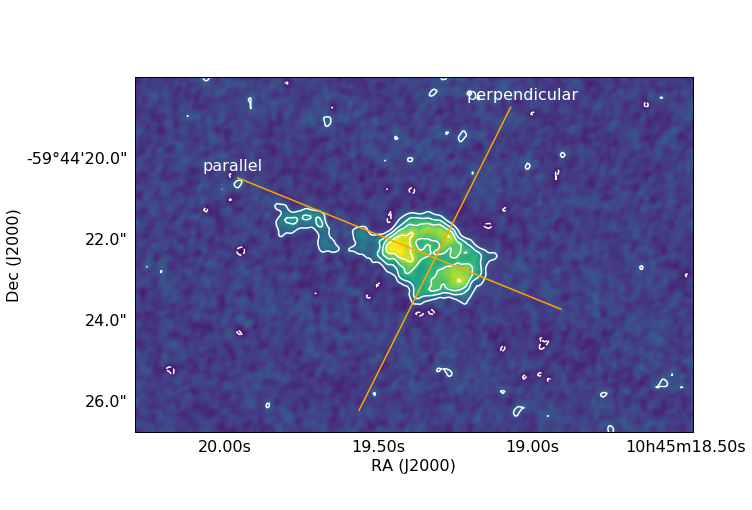} \\
\begin{array}{cc}
    \includegraphics[trim=0mm 60mm 0mm 0mm,angle=0,scale=0.475]{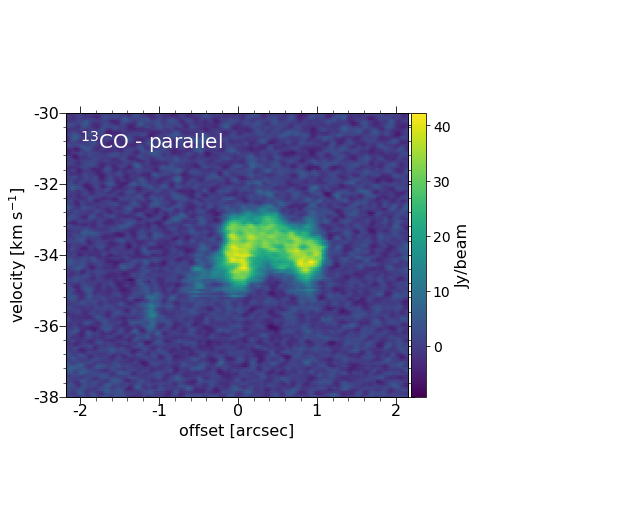} &
    \includegraphics[trim=50mm 60mm 0mm 0mm,angle=0,scale=0.475]{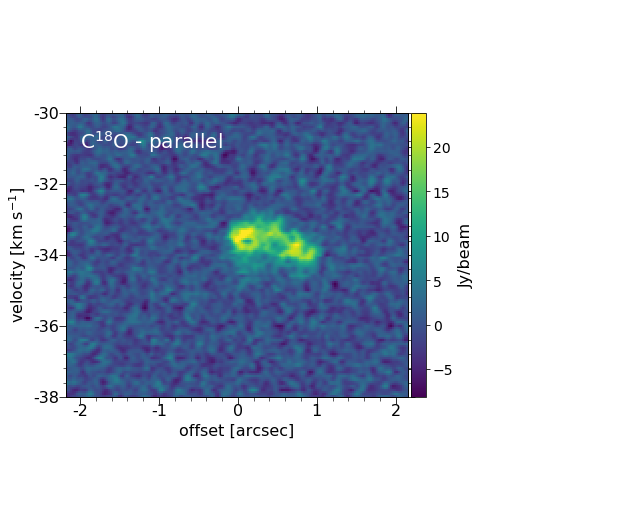} \\
    \includegraphics[trim=0mm 20mm 0mm 0mm,angle=0,scale=0.475]{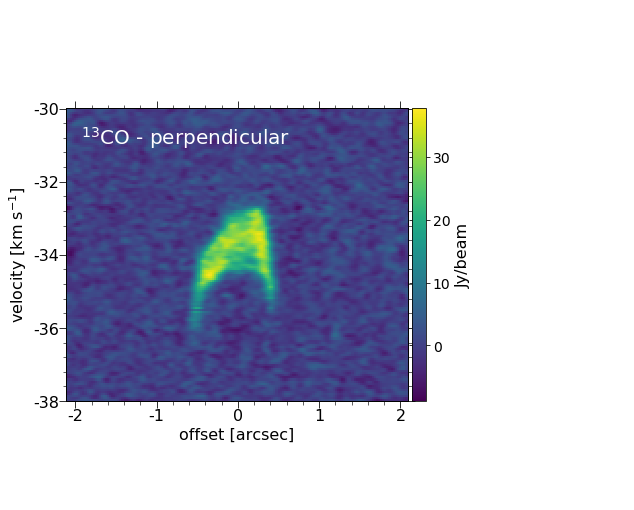} &
    \includegraphics[trim=50mm 20mm 0mm 0mm,angle=0,scale=0.475]{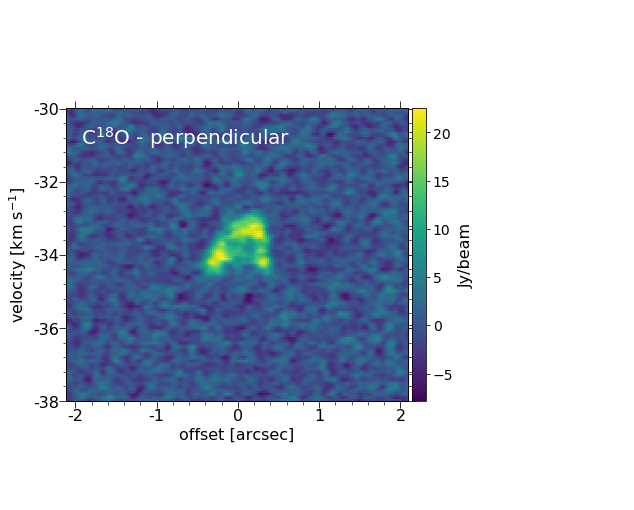}\\
    \end{array} \\
    \begin{array}{cccc}
    \includegraphics[trim=70mm 5mm 5mm 0mm,angle=0,scale=0.365]{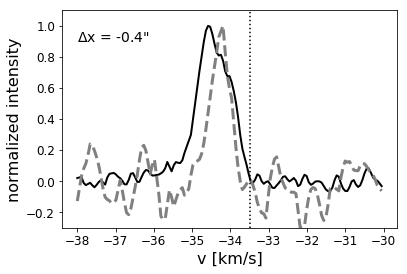} &
    \includegraphics[trim=10mm 5mm 5mm 0mm,angle=0,scale=0.365]{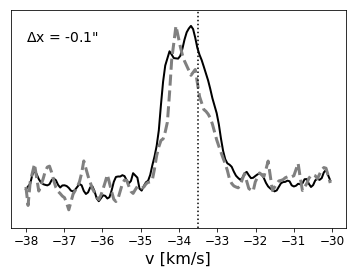} &
    \includegraphics[trim=10mm 5mm 5mm 0mm,angle=0,scale=0.365]{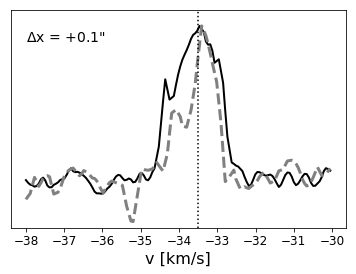} &
    \includegraphics[trim=10mm 5mm 5mm 0mm,angle=0,scale=0.365]{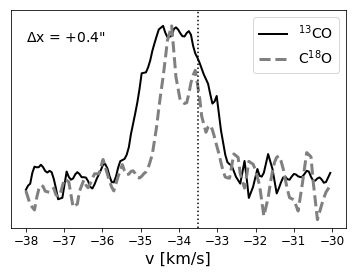} \\
    \end{array}
\end{array}$    
  \caption{
    \textit{Top:} Moment 0 map and contours showing the $^{13}$CO J=3-2 emission with lines showing the slices made through the globule to produce the P-V diagrams shown below. 
    Emission is summed in an aperture that is $0\farcs25$ wide. 
    \textit{Middle:} P-V diagrams of $^{13}$CO and C$^{18}$O J=3-2 through the major axis of the globule, parallel to the outflow axis  
    (\textit{upper})
    and perpendicular to the outflow axis
    (\textit{lower}). 
    \textit{Bottom:} Emission profiles as a function of velocity for a few positions in the perpendicular P-V diagrams (these correspond to the sum of emission in 5-pixel wide vertical slices through the P-V diagrams); $^{13}$CO is shown with a solid black line, C$^{18}$O with a gray dashed line, and a dotted line indicates the $v_{\rm LSR}$. 
}\label{fig:pv_globule} 
\end{figure*}

To examine the gas kinematics in the globule as a whole, we plot position-velocity (P-V) diagrams of emission parallel and perpendicular to the HH~900 outflow axis. 
Figure~\ref{fig:pv_globule} shows the slice locations and the P-V diagrams of $^{13}$CO and C$^{18}$O J=3-2 emission.
Both of the $^{13}$CO P-V diagrams trace a C-shaped velocity structure with velocities near the globule center that are $\gtrsim 1$~km~s$^{-1}$ redder than those near the edges. 
Most of the C$^{18}$O emission in the parallel P-V diagram is close to the $v_{\rm LSR}$, coinciding with the reddest velocities in the $^{13}$CO P-V diagrams. 
In contrast, the C$^{18}$O in the perpendicular P-V diagram traces a similar C-shape to that seen in the $^{13}$CO. 
We show a few vertical slices through the perpendicular P-V diagram in Figure~\ref{fig:pv_globule}. 
Emission profiles from both isotopologues tend to be quite broad in velocity space, with a peak intensity at bluer velocities closer to the edge of the globule. 
We discuss possible interpretations of this shape in Section~\ref{ss:glob_kin_int}.


\section{A photoevaporating jet+outflow system}\label{s:jet+outflow}
\begin{figure*}
  \centering
\includegraphics[trim=25mm 0mm 0mm 0mm,angle=0,scale=0.525]{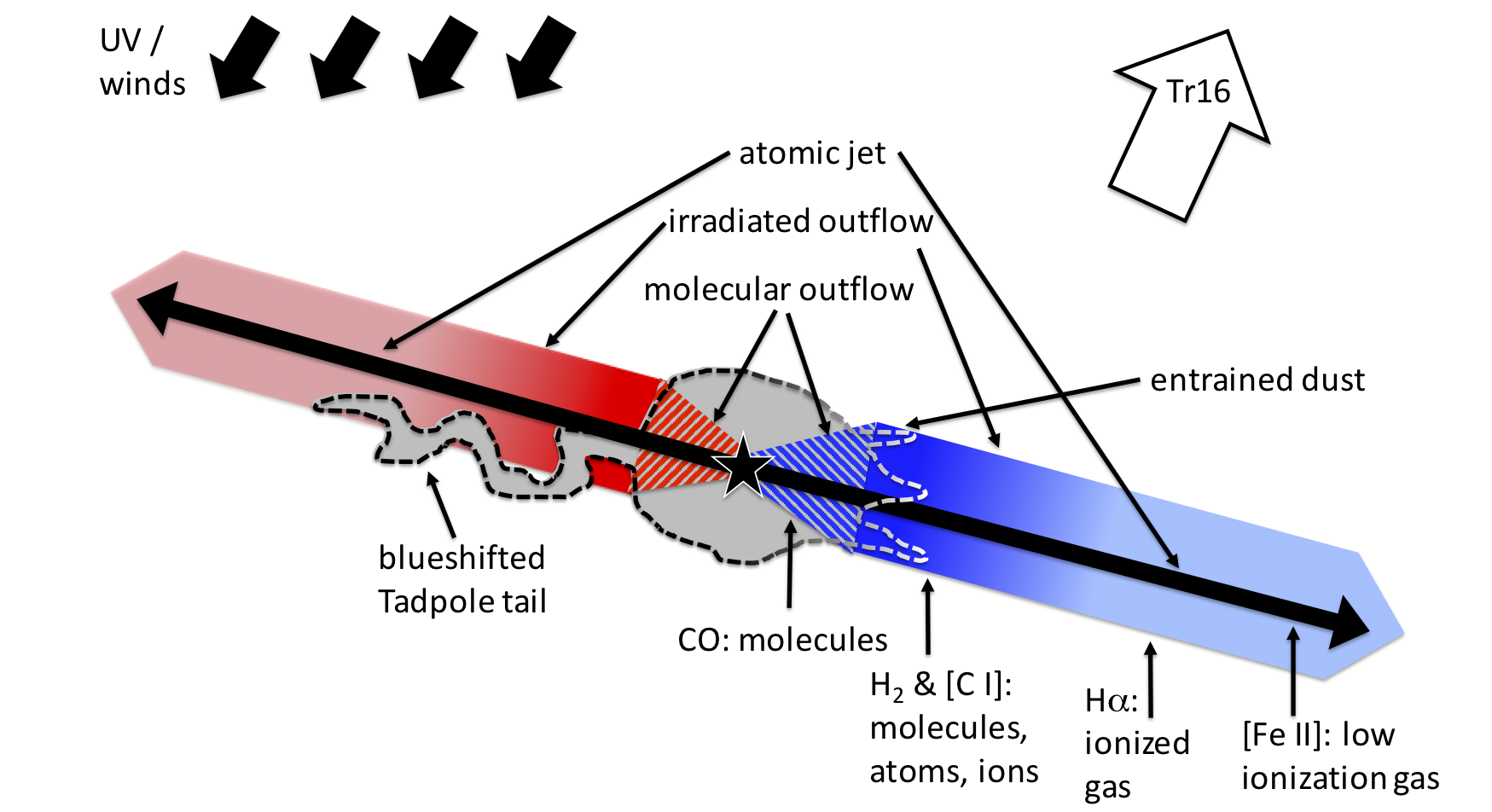}
  \caption{ 
  Cartoon rendering of the tadpole (gray) and the HH~900 jet+outflow system (red and blue).
    The \textit{jet} (shown in black) is seen in low-ionization species that trace the predominantly neutral gas; 
    the ionization state of the \textit{outflow} transitions from mostly molecular inside the globule to mostly ionized as it propagates into the H~{\sc ii} region (red and blue color gradients). 
    The tadpole tail is blueshifted in front of the redshifted side of the outflow and is physically distinct from the outflow.
  }\label{fig:hh900_cartoon} 
\end{figure*}

\citet{reiter2015_hh900} suggested that the HH~900 jet+outflow system is the irradiated analog of jet-driven molecular outflows seen in more quiescent environments \citep[e.g., HH~111,][]{lee2000,lef07}. 
Figure~\ref{fig:hh900_cartoon} shows a schematic of the HH~900 jet+outflow emerging from the tadpole. 
Only ionized gas tracers like H$\alpha$ trace the full extent of the outflow outside the globule. 
Hot molecular gas -- H$_2$ --
and 
partially ionized gas -- [C~{\sc i}] -- 
are detected in the irradiated outflow near the globule, but extend less than half the length of the irradiated outflow \citep[see][and Paper~I, respectively]{reiter2015_hh900}. With ALMA, we detect the molecular outflow inside the tadpole globule for the first time. 
The wide-angle molecular outflow (opening angle $\sim 50^{\circ}$) has a conical morphology that reaches the same width as the globule at the edge. 
This coincides perfectly with the H$\alpha$ outflow which is as wide as the globule where it emerges, before gradually tapering to terminus of jet (see Figure~\ref{fig:oflow_contours} and Paper~I). 
The abrupt end of the molecular outflow at the globule edge suggests that molecules are rapidly destroyed outside the protection of the optically thick globule. 
This structure is consistent with diagnostics presented in Paper~I that show an increase in the excitation in the outflow with increasing distance from the globule.

Multi-wavelength observations of the HH~900 jet+outflow system show layers of emission with the fastest material along the jet axis surrounded by slower gas in wider-angle components. 
This onion-like morphology and velocity structure has been seen in nearby outflow systems in less complicated environments \citep[e.g., DG~Tau, see][]{bac00,pyo03}. 
Forbidden emission lines (e.g., [Fe~{\sc ii}]) trace the highest density and most highly collimated portions of the jet. 
Molecular (H$_2$) and ionized (H$\alpha$) gas in the irradiated outflow are wider-angle, with the H$\alpha$ morphology coinciding with the jet ([Fe~{\sc ii}]) only at the terminus of the continuous inner outflow 
\citep{reiter2015_hh900}.  
The cold molecular outflow (CO) has a bi-conical morphology, originating from the \tyso\ in the center of the globule and opening to the same width as the H$\alpha$ and H$_2$ by the time it reaches the globule edge.

The velocities of each component are also consistent with an onion-like structure.
The fastest material,
with radial velocities $\pm \sim 40$~km~s$^{-1}$ traced by [Fe~{\sc ii}], 
is confined to the highly collimated jet. 
The surrounding layers of wider-angle gas are slower, reaching the same velocities as the fast jet only where the two components coincide at the terminus of the continuous inner outflow.  
The cold molecular gas traces the slowest material, with velocities roughly an order of magnitude slower than the jet-like components of HH~900 (see Section~\ref{ss:jet_props}).

Various models propose that molecular outflows result from jet entrainment \citep[e.g.,][]{rag93,lee2000,arce2001_hh300,ostriker2001} or represent a separate component altogether that is launched at slower velocities from larger radii in the disk \citep[e.g.,][]{pudritz1986,ferreira1997}. 
The tapering of H$\alpha$ emission with increasing distance from the tadpole is difficult to explain as there is no obvious physical mechanism that could recollimate the ionized outflow. 
However, the swept-up shell of a jet-driven molecular outflow will naturally have this morphology 
\citep[see Figure~2 in][]{arc07} 
and a Hubble-flow like velocity structure, as seen in H$\alpha$ \citep{reiter2015_hh900}.

To test the jet-driven outflow hypothesis, we compare the momentum flux of the atomic jet and the molecular outflow.
\citet{reiter2015_hh900} estimated the jet mass-loss rate to be 
$\dot{M}_{jet} \gtrsim 5 \times 10^{-6}$~M$_{\odot}$~yr$^{-1}$, within a factor of a few of the mass-loss rate we find in the molecular outflow, $\dot{M}_{out} \gtrsim 1 \times 10^{-6}$~M$_{\odot}$~yr$^{-1}$  (see Table~\ref{t:outflow_props}). 
The faster jet velocity ($\sim 100$~km~s$^{-1}$) compared to the molecular outflow ($\sim$few~km~s$^{-1}$) leads to a difference of more than an order of magnitude in the momentum flux from the two components.
If the jet mass-loss rate and therefore momentum are even higher, as suggested in Paper~I, then the jet can readily entrain the molecular outflow \citep[e.g.,][]{rag93,lee2000,arce2001_hh300,ostriker2001}.

External irradiation from nearby O-type stars illuminates the atomic jet, revealing material between shock fronts that would be too cold to see in more quiescent regions. 
This unique view makes HH~900 one of only a few sources where the momentum flux of both the jet and the outflow can be measured. 
In addition to testing whether the jet may drive the molecular outflow (as discussed above), systems like this are valuable to constrain the relative contribution of fast, collimated atomic jets and molecular outflows to the momentum budget in embedded star-forming regions.

Using a sample of jets and outflows in NGC~1333, \citet{dio17} found that the momentum flux of molecular outflows exceeds that of atomic jets. 
For the HH~900 jet+outflow system, as for a larger sample of externally irradiated jets in Carina \citep{reiter2017}, we find that atomic jets have comparable or larger momentum flux compared to molecular outflows.
\citet{reiter2017} determined the momentum flux in the atomic jet component of HH~900 
to be $\log(\dot{P}) \approx -2.7$, although  
evidence for higher densities presented in Paper~I suggest this may be even greater. 
In Section~\ref{ss:jet_props}, we computed a momentum flux in the cold molecular outflow of $\log(\dot{P}) \approx -4.1$ 
(taking the average of the red- and blue-shifted lobes), 
remarkably similar to the momentum flux \citet{reiter2017} estimate in the irradiated outflow
($\log(\dot{P}) \approx -4.3$). 
Different methods of measuring the atomic jet component may lead to differences in the fraction of the total mass included in the momentum flux estimate \citep[see][]{reiter2017}.


\section{Cold molecular gas and dust in the tadpole-shaped globule}\label{s:discussion_intro}

The ALMA data presented in this paper provide the first detection of the deeply embedded \tyso. 
Both \citet{reiter2015_hh900} and Paper~I argue that high densities in the tadpole globule provide several magnitudes of extinction to obscure the \tyso. 
We can estimate the extinction to the \tyso\ from the 
molecular column density using 
the relationship between $A_V$ and N($^{13}$CO) derived by \citet{dickman1978}, 
\begin{equation}
A_v \approx (4.0 \pm 2.0) \times 10^{-16} \times \mathrm{N(^{13}CO)} \,\,\mathrm{mag}
\end{equation}
assuming local thermodynamic equilibrium (LTE). 
Using the median optical-depth corrected $^{13}$CO column density, 
N($^{13}$CO) $\sim 1.25 \times 10^{17}$~cm$^{-2}$,  we find $A_V \sim 50$~mag. 
This high $A_V$ is consistent with a protostar too embedded to be seen at shorter wavelengths, but more than an order of magnitude higher than the $A_V \sim 2.5$~mag derived from optical hydrogen recombination lines in Paper~I. 
However, the optical diagnostics only probe the extinction in front of the ionized layer on the globule surface. 
The higher $A_V$ estimate from the N($^{13}$CO) includes the extinction from gas and dust in the tadpole globule itself, and thus reflects a much larger column of material.

The temperature structure of the globule shows evidence of the external influence of the environment (see Figure~\ref{fig:bTemp} and Section~\ref{s:tadpole_temp}). 
Multiple lines of evidence point to a positive radial temperature gradient with
hotter gas near the surface of the globule compared to the interior (see Figure~\ref{fig:bTemp} and Section~\ref{s:tadpole_temp}). 
While the surface of the globule is hot and ionized (see Paper~I),  
dust deep in the interior of the tadpole remains cold, with 
continuum emission better fit with models with lower temperatures ($T<30$~K, see Figure~\ref{fig:central_SED} and Section~\ref{ss:dust}). 
We serendipitously detect DCN J=3-2 in the center of the globule, which we take as evidence that the gas is also cold (see Section~\ref{ss:circumstellar}). 
Chemical reactions in cold gas can enhance the abundance of deuterated species, especially in high density gas where CO is depleted \citep[e.g.,][]{dalgarno1984,millar1989,turner2001}.
Where DCN has been detected in hot sources \citep[e.g., Orion KL,][]{mangum1991}, it is thought to have been returned to the gas phase from icy grain mantles only recently. 
Using the estimated dust mass surrounding the \tyso\ ($\sim 1$~\Msun, see Section~\ref{ss:dust}) and assuming a uniform density profile, the average density in the globule is $\gtrsim 10^6$~cm$^{-3}$. 
This is well above the density where gas and dust are expected to be thermally coupled \citep[see, e.g.,][]{goldsmith2001,galli2002}.
Together, this suggests cooler temperatures deep in the tadpole globule, where it is shielded from the harsh environment by the large column density of material.

\subsection{Globule survival}\label{ss:glob_mass} 

In Paper~I, 
we estimated that the tadpole is being photoevaporated at a rate $\dot{M} \sim 5 \times 10^{-7}$~M$_{\odot}$~yr$^{-1}$.
Assuming a constant photoevaporation rate, we can compute the remaining globule lifetime. 
We have estimated the mass in two different but complementary ways: 
(1) using N(C$^{18}$O), assuming it is optically thin; and 
(2) from the dust continuum (see Section~\ref{ss:dust}). 
Using N(C$^{18}$O), we estimate the mass of molecular gas in the tadpole globule to be $\sim 0.6$~M$_{\odot}$ (see Section~\ref{ss:alma_masses}). 
Given the high optical depths in the tadpole, it is likely that C$^{18}$O is also optically thick, so this mass estimate is a lower limit. 
Continuum emission is the most optically thin diagnostic available. 
From the dust continuum, we estimate a mass 
$M_{\mathrm{tadpole}} \approx 1.9$~M$_{\odot}$, 
a factor of $\sim 3$ higher than the estimate from C$^{18}$O ($\sim 0.6$~\Msun) 
and roughly half the Bonnor-Ebert mass estimated in Paper~I ($\sim3.7$~\Msun). 
We note that the C$^{18}$O and dust continuum trace different spatial distributions (see Figures~\ref{fig:alma_data} and \ref{fig:cont}), so the sum of the two masses may be a better reflection of the mass of the globule. 
In the following, we focus on the mass estimate from the dust continuum of the tadpole as a whole. 

Assuming a constant photoevaporation rate, a 1.9~\Msun\ globule will be completely ablated in $\sim 4$~Myr. 
The remaining globule lifetime will be 
shorter ($\sim 1$~Myr) or 
longer ($\sim 7$~Myr, see Paper~I)
for the lower and higher mass estimates, respectively. 
For these remaining globule lifetimes, the \tyso\ will emerge into the H~{\sc ii} region as the high-mass stars in nearby Tr16 explode as supernovae. 

Fossil evidence in the Solar System meteorites requires that the Sun formed near at least one dying high-mass star \citep[e.g.,][]{adams2010}, suggesting that the tadpole provides an interesting environment for planet formation. 
 Although the planet formation process is still not fully understood, there is growing evidence that it happens early. For example the concentric rings of HL~Tau with an age of $\sim 1$~Myr  \citep{alma2015}
 can be explained in terms of three embedded planets (\citealt{dipierro2015}, although see \citealt{zhang2015}). 
 The pebble accretion driven model of planet formation in the famous 7-planet hosting Trappist-1 system also operates on a $\sim 1$~Myr timescale \citep{ormel2017,schoonenberg2019}. 
 Given that photoevaporation of planet-forming disks by other stars in a cluster can severely constrain planet formation even if planet formation happens early 
\citep[e.g.,][]{haworth2018,winter2018,concha2019,nicholson2019} 
any progress towards planet formation in an embedded stage such as within a globule is very important.

Model fits to the continuum spectral index (see Section~\ref{ss:dust}) suggest that grain growth is already underway in the tadpole. 
High optical depths and high densities in the globule shield the disk from the harsh radiative environment (as suggested by the positive radial temperature gradient, see Section~\ref{s:tadpole_temp}). 
The inferred globule lifetime of $\sim 4$~Myr could in principle permit the entire planet-formation process to take place while shielded from external irradiation from the nearby stellar cluster. 
Even the shortest estimated globule lifetime of $\sim 1$~Myr (which we consider to be an underestimate) would allow significant progress toward planet formation, for example through rapid grain growth and radial drift into small radii \citep[e.g.][]{2012A&A...539A.148B},   
before the globule is completely ablated. 
As discussed in the previous paragraph, discs around young ($\sim1$\,Myr) sources show rings and other characteristics thought to be due to planet formation \citep{alma2015}.
In this case, dense star-forming globules may be an important class of planet-forming systems in stellar clusters.

Disk evolution within this context may still differ from local clouds given the high optical depths and high densities in the tadpole. 
For example, if external feedback acts to accelerate the collapse of the globule, 
the disk accretion rate may also be enhanced, altering the dynamics in the planet-forming disk. 
It is unclear whether or how such star-forming globules may be enriched with the short-lived radioactive isotopes that play an important role in the geochemical evolution of terrestrial exoplanets \citep[e.g.,][]{grimm1993}. 
Recent studies point to pre-supernova mass-loss as an important source of short-lived elements \citep{lugaro2018} that may also provide an earlier enrichment pathway. 
Indeed, the current abundance of key elements like $^{26}$Al in Carina appears to be on the order of the value inferred for the early Solar System \citep[see discussion in][]{reiter2019}.

\subsection{Globule kinematics}\label{ss:glob_kin_int}

The ALMA data presented in this paper reveal the deeply embedded \tyso\ for the first time, providing unambiguous evidence for star formation in the opaque tadpole globule.
Infalling gas often shows complex self-absorbed line profiles \citep[e.g.,][]{walker1986,tafalla1998,narayanan2002,reiter2011_kinematics}, 
however line profiles from the tadpole tend to be single-peaked (see Figure~\ref{fig:modelCO}). 
Bright-rimmed clouds that have been affected by external feedback often do not show the asymmetric line profiles characteristic of infall 
\citep[e.g.,][]{devries2002,thompson2004}. 
This absence has been attributed to other dynamical effects like rotation, pulsation, or a combination of collapse and expansion \citep[e.g.,][]{redman2004,keto2006,gao2010,wang2012}. 
In the case of the tadpole, 
the positive radial thermal profile can explain the absence of characteristic line asymmetries:  
hot intervening material does not absorb emission from the colder inner regions, and emission closer to the line peak traces warmer, more external gas (see Section~\ref{s:tadpole_temp}).

Position-velocity slices through the globule trace a C-shaped morphology (see Figure~\ref{fig:pv_globule} and Section~\ref{ss:gas_kin}). 
Tracers like C$^{18}$O probe the kinematics of the colder gas deeper in the globule than optically thick lines like $^{13}$CO. 
We expect that global motions will produce a similar morphology in the C$^{18}$O P-V diagrams. 
While the shape of the emission in both lines is similar, it is not clear what kinematic structure this traces. 
C-shaped features have been seen in P-V profiles of ionized gas tracing much larger scales 
\citep[e.g.,][]{keto2002,keto_wood2006} and interpreted as infall. 
By fitting the profile for infall and rotation, \citet{keto2002} derived an infall velocity of $\sim 4.5$~km~s$^{-1}$. 
We do not see evidence for global rotation of the tadpole.

We estimate the free-fall velocity in the tadpole, $v_{\mathrm{ff}} = \sqrt{2GM/r}$.
For the dust mass computed in Section~\ref{ss:dust} ($\sim 1.9$~\Msun), we find $v_{\mathrm{ff}} \sim 1.2$~km~s$^{-1}$ 
in the absence of any pressure support. 
This is similar to the velocity difference we observe between the center and the edge of the tadpole (see Figure~\ref{fig:pv_globule}). 
The estimated free-fall velocity will be lower ($v_{ff} \sim 0.7$~\kms) for the mass estimated from the C$^{18}$O ($0.6$~\Msun) or higher ($v_{ff} \sim 1.6$~\kms) for the Bonnor-Ebert mass derived in Paper~I ($\sim 3.7$~\Msun).  
\citet{mottram2013} found infall velocities (extrapolated to 1000~AU) of $\sim 1$~km~s$^{-1}$ from radiative transfer modeling of water emission lines seen in low-mass protostars, assuming a free-fall velocity profile. 
Other authors using similar methods on different lines and higher spatial resolution data \citep[e.g.,][]{difrancesco2001} find somewhat smaller velocities.

On the theoretical side, few studies exist that present spatially-resolved gas kinematics. 
\citet{haworth2013} simulated bright-rimmed clouds
compressed by an external ionizing source, and produced
synthetic observations of the J=2-1 transition of $^{12}$CO,
$^{13}$CO, and C$^{18}$O. The simulated data were optimised for comparison with observations from the James Clerk Maxwell Telescope, a 15~m single-dish facility with a beamsize of $\sim 22"$ at these frequencies, more than an order of magnitude larger than the synthesized beamsize of our ALMA data. As a result, example clouds presented in that
paper are $\sim 50\times$ the size of the tadpole.

\section{Comparison with similar objects}
\subsection{Other protostars in Carina}\label{ss:ysos}
Two flattened, disk-like structures have been reported in Carina, both from protostars embedded in small globules like the tadpole \citep{mesa-delgado2016}. 
Both disks are small, marginally resolved with a $0\farcs03 \times 0\farcs02$ beam, yielding approximate disk radii of $\sim 60$~AU.
A disk of similar size around the \tyso\ would be unresolved in our ALMA data.  

The 229~GHz flux densities of the two disk detections in \citet{mesa-delgado2016} are 0.9~mJy and 1.5~mJy, 
a factor of $\sim 4$ lower than the flux density of the \tyso\ embedded in the tadpole. 
The HH~900 jet mass-loss rate is an order of magnitude higher than that of either of jets driven by the protostars in the \citet{mesa-delgado2016} study, suggesting that the higher flux might come from a more massive circumstellar disk.

\subsection{Other globules}

Globules are seen in many H~{\sc ii} regions and have been studied at shorter wavelengths for decades \citep[e.g.,][]{bok1948,herbig1974,reipurth1983}. 
Several authors also note a tadpole-like morphology \citep[e.g.,][]{brandner2000,sahai2012_tadpole,wright2012}. 
Tails of ionized gas coming from these objects typically point away from the ionizing sources. 
Their morphology resembles the tear-drop shape of the nebulosity surrounding true proplyds \citep[e.g.,][]{odell1994,johnstone1998}. 
Indeed, the tadpole was originally identified as a candidate proplyd \citep{smith2003}, despite what looked like two separate and oppositely-directed tails. 
These tails have now been identified as the HH~900 jet+outflow system. 
Molecular line observations of the tadpole and other globules demonstrate that their masses are orders of magnitude higher than typical proplyds  \citep[e.g.,][respectively]{sahai2012_tadpole,sahai2012,mann2010,mann2014}.

Like other globules seen in H~{\sc ii} regions, the tadpole is primarily seen in silhouette \citep{smith2010}. 
Paper~I summarizes the morphological clues that suggest that the globule lies in front of Tr16: 
(1) its $v_{LSR}$ is blueshifted by $\sim 10$~\kms\ compared to the $v_{LSR}$ of Carina \citep[see Section~\ref{s:results} and][]{rebolledo2016}; 
(2) the tadpole tail is further blueshifted with respect to the head (see Figure~\ref{fig:alma_data}); 
(3) the ionization front on the globule surface is bright on both the side closest to and further from Tr16 (see Paper~I); and 
(4) the opaque globule center suggests minimal  illumination on the
near side.
If the system does lie in front of Tr16, this may help explain the absence of a spatially-resolved tail of ionized gas seen extending from the tadpole in the direction away from Tr16.

Globules may instead be the remnants of dust pillars where the high-density material at the head has separated from the more diffuse body \citep[e.g.,][]{hester1996,ercolano2011}. 
The tadpole may be one such pillar remnant that pointed toward Tr16. 
We observe a velocity gradient in the molecular emission along the tadpole, with material in the tail a few km~s$^{-1}$ bluer than in the head. 
Motion in the tail is not affected by the molecular outflow, which is redshifted on this side of the \tyso\ (see Figures~\ref{fig:oflow_contours}, \ref{fig:hh900_cartoon}, and \ref{fig:pv_globule}). 
The peculiar tail may be the remnants of high-density pillar spine, as seen in a few other pillars in Carina \citep[e.g.,][]{klaassen2020}.

\citet{gahm2007} identified many small globulettes in the Rosette Nebula with typical sizes $\sim 2500$~AU, similar to the observed size of the tadpole. 
Using the same extinction estimate technique, \citet{grenman2014} identified hundreds of globulettes in Carina, including the tadpole. 
Most observations of  the molecular gas content of these small globules have been performed using single-dish facilities
with beamsizes $\sim 18^{\prime\prime}$--$27^{\prime\prime}$, an order of magnitude larger than the typical globulette size.  
Nevertheless, some trends have been found in the properties of the large globulettes observed in this way. 
Both \citet{gahm2013} and 
\citet{haikala2017}, targeting the Rosette and Carina, respectively, 
find systematically higher masses than estimates from the extinction. 
\citet{haikala2017} confirm that the globulettes in Carina are smaller, higher density, and tend to have higher linewidths than those in the Rosette.

Globules that show evidence for an embedded protostar provide the best comparison with the tadpole. 
Both \citet{sahai2012} and \citet{haikala2017} observed another star-forming globule in Carina that contains the HH~1006 jet. 
\citet{sahai2012} demonstrated that the mass of molecular gas is too high ($\sim 350$~M$_{Jup}$) for this object to be a proplyd. 
Extended linewings probably trace the molecular outflow associated with the HH~1006 jet \citep{reiter2016}. 
\citet{haikala2017} also 
find that the linewidths in the HH~1006 globule are larger than those in globulettes without evidence for star formation. 
Like the HH~1006 globule, we find that the tadpole is more massive than previous estimates ($\sim 1.9$~M$_{\odot}$, see Section~\ref{ss:dust}). 
Broad linewidths, especially in $^{12}$CO, reveal the molecular outflow associated with the HH~900 jet+outflow system (see Figure~\ref{fig:oflow_profile}). 
The tadpole is too opaque to probe the circumstellar disk, unlike HH~1006 where
\citet{mesa-delgado2016} estimated a disk mass of $\sim 27$~M$_{Jup}$ (assuming T=40~K) around the embedded protostar. 

More detailed comparisons will be possible in the future as ALMA observes the internal structure of more of the globules in Carina and other H~{\sc ii} regions.

\section{Conclusions}\label{s:conclusions}

We present spatially and spectrally resolved ALMA observations of a small globule, the tadpole, in the Carina Nebula. 
Previous observations at shorter wavelengths of the HH~900 jet+outflow system that emerges from the globule suggested a protostar and molecular outflow hidden in the opaque globule \citep[see][and Paper~I]{reiter2015_hh900}. 
Our ALMA data have angular resolution comparable to \emph{HST}, allowing us to 
conduct a detailed analysis of the embedded protostar, molecular outflow, and tadpole globule as a whole. Our main conclusions are as follows: 

\begin{itemize}
    \item We detect the molecular outflow associated with the HH~900 jet+outflow system for the first time in $^{12}$CO J=2-1. The wide-angle molecular outflow traces a biconical shape that joins smoothly with irradiated outflow seen outside the globule (Paper~I). 
    \item The momentum flux in the HH~900 molecular outflow is comparable to that in the atomic jet, consistent with the jet-driven outflow morphology seen in the overall system.  
    \item The \tyso\ is detected on the jet axis inferred from optical and near-IR images, and is the clear origin of the associated molecular outflow. Continuum emission near the \tyso\ is marginally resolved, and appears to be slightly flattened perpendicular to the outflow axis. Optical depths in the globule are too high to trace the kinematics of the gas closest to the source. The serendipitous detection of DCN J=3-2 surrounding the \tyso\ suggests that gas in the center of the globule remains cold, despite evidence that the harsh environment heats the globule surface to higher temperatures. 
    The best-fit continuum spectral index of $\sim 2$ suggests that grain growth is already underway in the \tyso.
    \item We measure high optical depths in the CO isotopologues, in line with previous estimates of high densities in the globule that render it opaque at shorter wavelengths. 
    \item From the dust continuum emission, we estimate a globule mass $\sim 1.9$~M$_{\odot}$.
    \item For the photoevaporation rate estimated in Paper~I, this suggests that the remaining lifetime of the tadpole is $\sim 4$~Myr. While external photoevaporation rapidly destroys protoplanetary disks around exposed stars, this long globule lifetime suggests that the \tyso\ will remain shielded from the environment for much of the planet-formation timescale. Evidence for grain growth in the tadpole suggests that dense globules may be an important class of planet-forming systems in stellar clusters.  
    \item Position-velocity slices through the tadpole (perpendicular to the outflow) show a C-shape morphology in $^{13}$CO and C$^{18}$O.  This emission structure has been interpreted as infall when observed in other sources. Unlike colder sources where characteristic line asymmetries indicate infall, line profiles in the tadpole are remarkably symmetric.  Whether gas kinematics in the tadpole are consistent with infall remains unclear. 
\end{itemize}

The \tyso\ and jet+outflow system are clearly much younger than nearby Tr16. 
Radiatively-driven implosion may be an essential element of the dynamical evolution of the globule.
A more thorough discussion of how environment affects the kinematics and evolution of the globule will be presented in Paper~III.

\section*{Acknowledgements}
M.R. would like to thank Ted Bergin, John Bally,  Libby Jones, and Carolyn Atkins for helpful discussions. 
M.R. was partially supported by a McLaughlin Fellowship at the University of Michigan and 
has received funding from the European Union's Horizon 2020 research and innovation programme under the Marie Sk\'{l}odowska-Curie grant agreement No. 665593 awarded to the Science and Technology Facilities Council. 
T.J.H is funded by a Royal Society Dorothy Hodgkin Fellowship.
A.F.M. is funded by a NASA Hubble Fellowship. 
G.G. acknowledges support from CONICYT project AFB-170002. 
This paper makes use of the following ALMA data: ADS/JAO.ALMA\#2016.1.01537.S. ALMA is a partnership of ESO (representing its member states), NSF (USA) and NINS (Japan), together with NRC (Canada) and NSC and ASIAA (Taiwan) and KASI (Republic of Korea), in cooperation with the Republic of Chile. The Joint ALMA Observatory is operated by ESO, AUI/NRAO and NAOJ.
This work uses observations made with the NASA/ESA Hubble Space Telescope, obtained from the Data Archive at the Space Telescope Science Institute, which is operated by the Association of Universities for Research in Astronomy, Inc., under NASA contract NAS 5-26555.
The HST observations are associated with GO 13390. 
This research made use of Astropy,\footnote{http://www.astropy.org} a community-developed core Python package for Astronomy \citep{astropy:2013, astropy:2018}.
This research made use of APLpy, an open-source plotting package for Python \citep{robitaille2012}. 
This work has made use of data from the European Space Agency (ESA) mission
{\it Gaia} (\url{https://www.cosmos.esa.int/gaia}), processed by the {\it Gaia}
Data Processing and Analysis Consortium (DPAC,
\url{https://www.cosmos.esa.int/web/gaia/dpac/consortium}). Funding for the DPAC
has been provided by national institutions, in particular the institutions
participating in the {\it Gaia} Multilateral Agreement.




\bibliographystyle{mnras}
\bibliography{bibliography_mrr} 



\appendix

\section{Coordinate correction of ACS/HST image}\label{s:astrometry}
\begin{figure*}
  \centering
    (a)\includegraphics[width=0.45\textwidth]{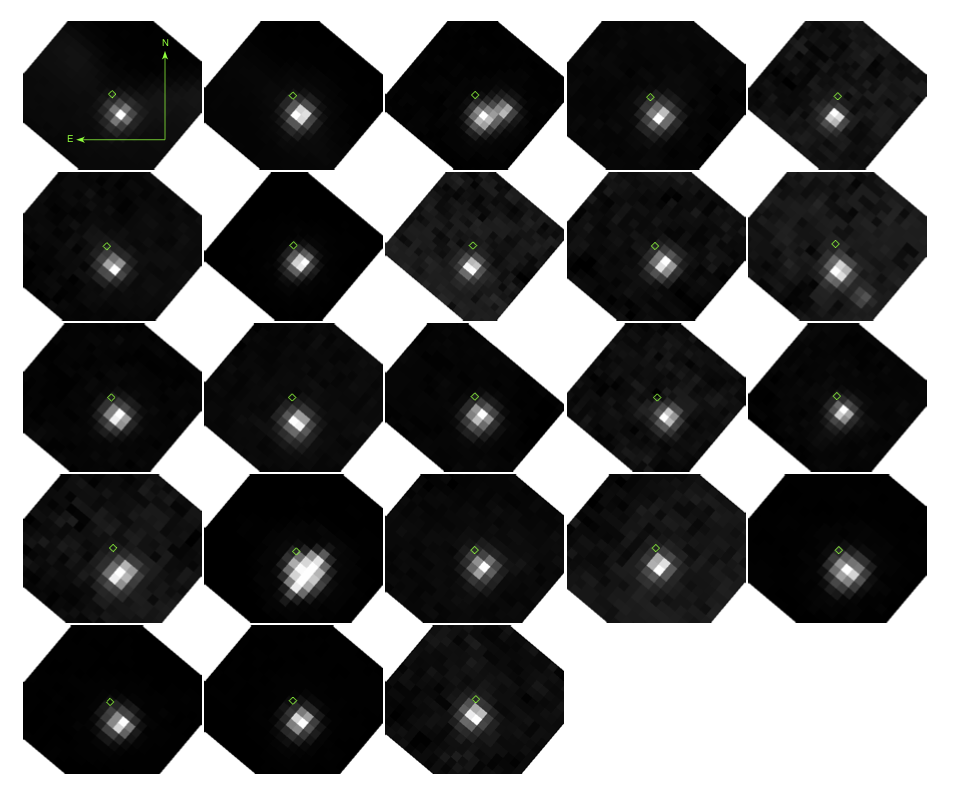}%
    (b)\includegraphics[width=0.45\textwidth]{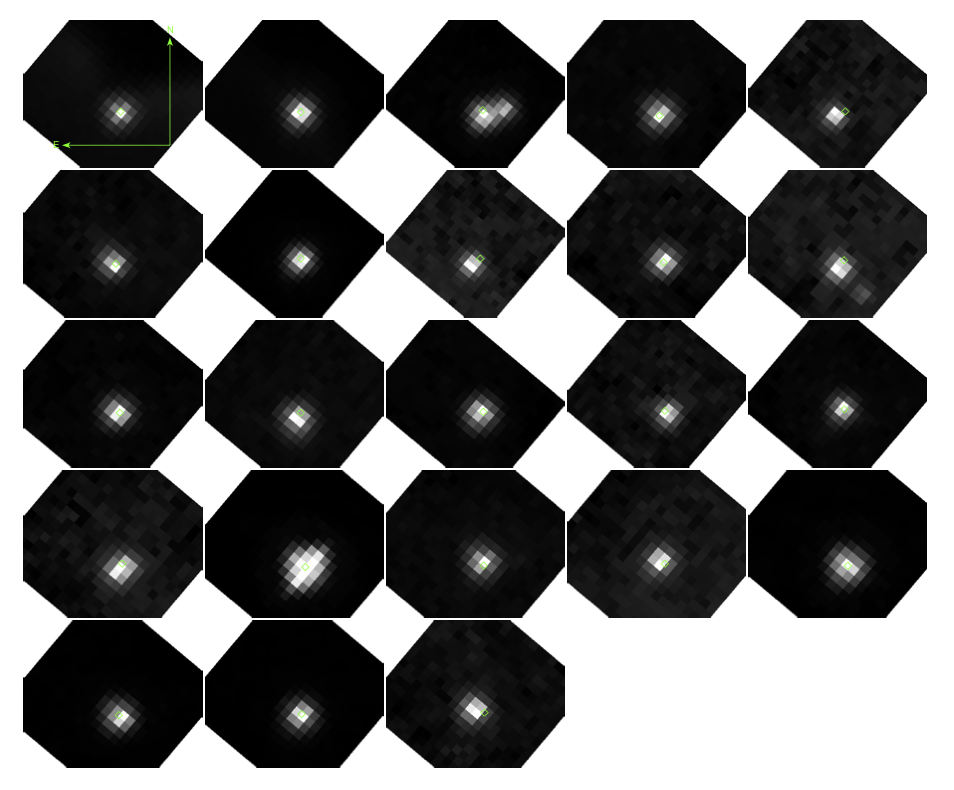}%
    \caption{HST/ACS H$\alpha$ image around Gaia sources in the field. Panel (a) shows Gaia sources' positions as a green diamonds compared with the original ACS/HST image. Panel (b) shows the same but onto the image with corrected coordinates.\label{fig:coorshift_comparison}}
\end{figure*}

For this work, we corrected the coordinates of the ACS/HST H$\alpha$ (F658N filter) image taken 18 of July 2005 \cite{reiter2015_hh900} by comparing this image with the position of 23 Gaia  sources in the field \citep{gaia2018,gaia2016}. Figure \ref{fig:coorshift_comparison}
shows these sources and the position of each associated Gaia source  
by the time of the ACS/HST observations according to their proper motion. Panel (a) marks the position of the Gaia coordinates in the original H$\alpha$ image. It is apparent that the HST coordinate solution is displaced on average by about $(\Delta \alpha,\Delta \delta)=(-0\farcs12,-0\farcs05)$, determined  by comparing the peak position of the HST sources with the Gaia coordinates. Panel (b) shows  H$\alpha$ image with this correction applied.
The centered H$\alpha$ image  shown in Figure \ref{fig:landscape} is used to correct the [Fe~{\sc ii}] image in Figure \ref{fig:oflow_contours}.
We consider this simple coordinate correction sufficiently accurate for the purpose of the present study.

\section{Maximum recoverable scale. Short-baseline filtering.}\label{s:filtering}
\begin{figure}
  \centering%
  \includegraphics[height=0.5\textwidth,angle=-90]{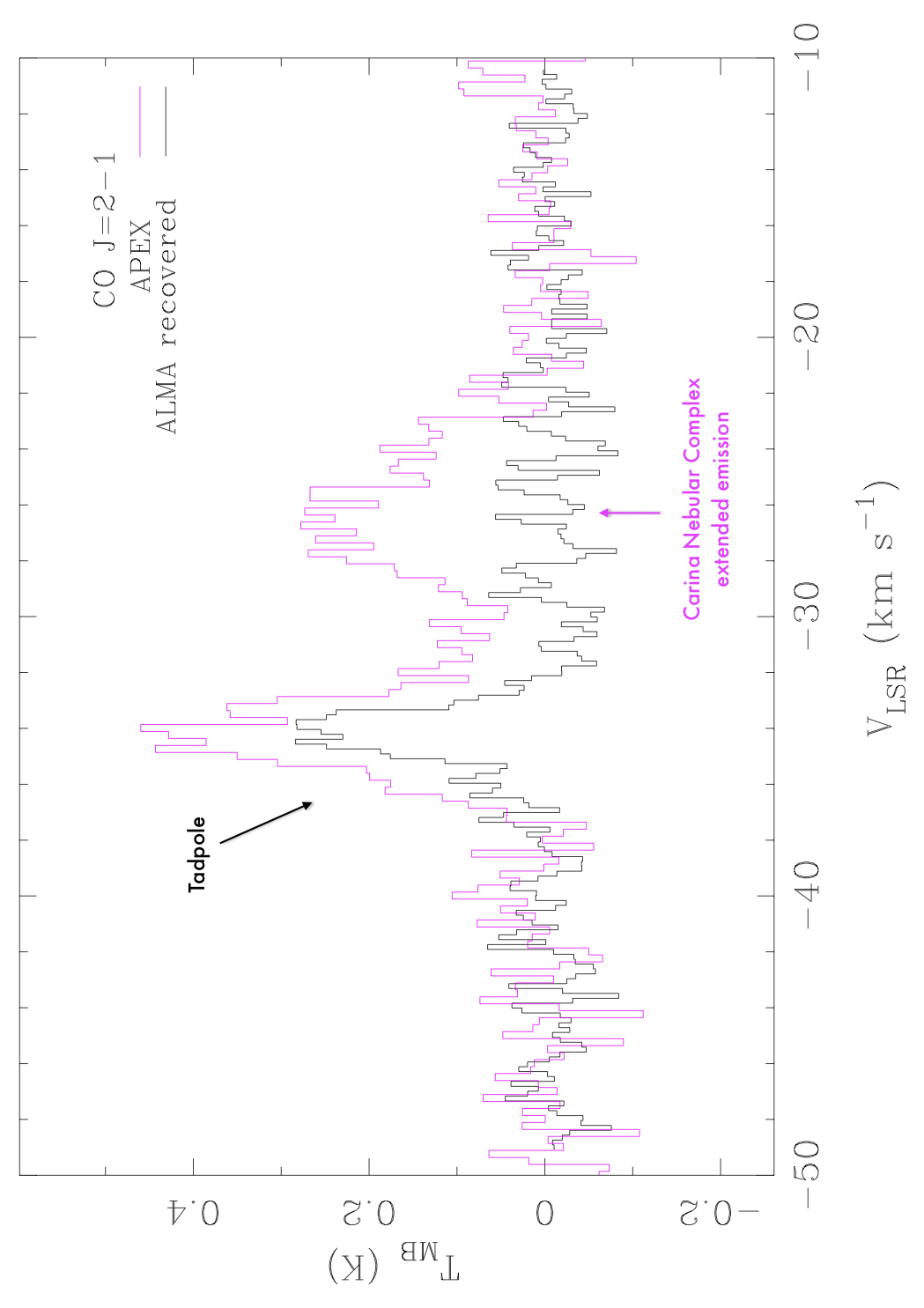}
  \caption{Comparison between the main beam corrected spectrum of the CO J=2-1 line toward HH 900 taken using APEX and 
  the emission recovered by ALMA.\label{fig:compar_ALMA_APEX}}
\end{figure}

We evaluate how much ALMA recovers the large scale emission from the tadpole. For that, we compare the ALMA data with the emission measured  with the APEX telescope (project O-097.F-9337A.2016) 
of the CO J=2-1 toward HH~900. Position switching spectra was taken  toward the tadpole position on 04 April 2016, 
integrating on-source $\approx24$ min, allowing us to reach an rms of 0.04 K main-beam corrected temperature ($\eta_{\rm MB}=0.7$). The main beam FWHM of APEX at the CO J=2-1 frequency is $26\farcs3$. 
We weight the ALMA primary-beam corrected CO J=2-1 line cube (natural weighting) with a 2D Gaussian  of FWHM 
equivalent to the APEX main beam size. This results in the black spectrum shown in Figure \ref{fig:compar_ALMA_APEX}.  This spectrum is  very similar to that obtained just by  integrating the entire  ALMA cube because the APEX telescope has the same diameter as the  ALMA main array antennae.

The APEX spectrum in Figure \ref{fig:compar_ALMA_APEX} shows two main components. There is a $3.2$ \kms\ wide component centered at $-34.2$ \kms\ and likely associated with the tadpole.  A second component is   $5.7$ \kms\ wide and it is  centered at $-26.5$ \kms. This second component is probably associated with the Carina Nebular Complex large scale emission \citep{rebolledo2016,klaassen2020}.   

Figure \ref{fig:compar_ALMA_APEX} indicates that our ALMA observations recover approximately a 70\% from the tadpole emission. If this difference is real (in contrast to being artificially produced by, for example, calibration problems)
and  is due to interferometer short baseline filtering, then the filtered emission should consist of an extended diffuse molecular envelope encompassing the tadpole. 
As expected, our ALMA observations cannot recover the widespread Carina Nebular emission from the second component. 

\section{PCYC~838}\label{s:pcyc838}

We detect continuum emission and CO from the YSO that lies in western limb of the outflow, PCYC~838 (see Figures~\ref{fig:landscape} and \ref{fig:alma_data}). 
We do not detect $^{13}$CO at the location of PCYC~838, so we assume that the $^{12}$CO emission is optically thin. 
We use this to compute a median log(N(CO)$_{thin}$)=15.9 for PCYC~838, significantly lower than that estimated for the tadpole globule (see Table~\ref{t:molecular_props}). 

Table~\ref{t:fluxes_continuum} gives the flux densities within a radius of 0\farcs3 ($r_0=690$~AU) of PCYC~838 and the spectral index of a power-law fit.
The mass derived from the continuum fluxes (Table~\ref{t:fluxes_continuum}) toward PCYC 838 using the same hypotheses as in Section~\ref{ss:dust} is about $0.2$ \Msun, assuming 15~K. The stellar mass estimated by \citet{povich2011}  is $\sim 2.5$~\Msun\ based on model fits to the spectral energy distribution. 
Because this source is visible at optical wavelengths, it is likely less embedded and it may be associated with a higher dust temperature. Assuming $T_{\rm dust}=50$ K, we obtain a mass of $0.05$ \Msun. 
The spectral index of the continuum emission toward PCYC~838 is also low, suggesting dust coagulation in this source, as in the \tyso. 
Unlike the \tyso, PCYC~838 is unresolved at all wavelengths.  
\begin{table*}
\caption{Position and flux densities of PCYC~838.\label{t:pcyc838_fluxes}}
\centering
\begin{footnotesize}
\begin{tabular}{llllllll}
\hline\hline
Source & R.A. & decl.&  217.1~GHz & 232.2~GHz &  331.6~GHz & 343.0~GHz & $\alpha^{\dagger}$\\
        & (J2000) & (J2000) & [mJy] & [mJy] & [mJy] & [mJy] & \\
        \hline
PCYC~838        &10:45:18.798 & $-$59:44:23.78 & 1.21 & 1.44 &  2.64 &  2.73 & $1.8\pm0.5$\\\hline
\multicolumn{8}{l}{$^{\dagger}$Spectral index of best power-law fit to flux densities ($S_\nu\propto\nu^\alpha$).}\\
\end{tabular}
\end{footnotesize}
\end{table*}

\section{Optical depth maps}\label{s:tau_maps}

\begin{figure}
  \centering%
  $\begin{array}{c}
    \includegraphics[trim=10mm 50mm 0mm 50mm,angle=0,scale=0.35]{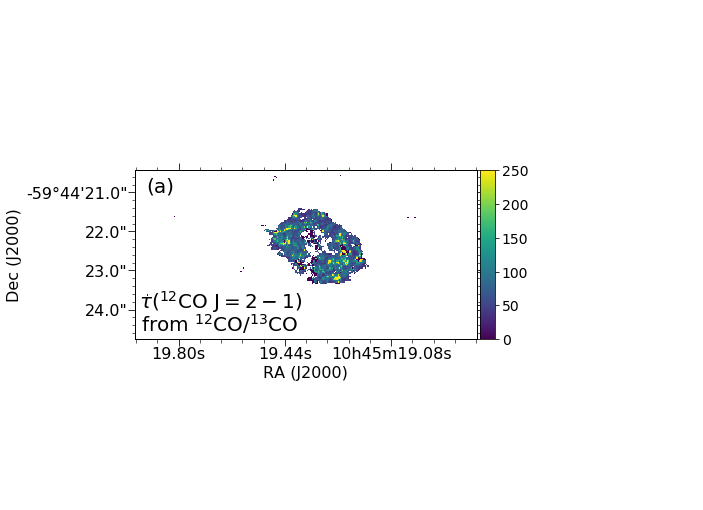} \\
    \includegraphics[trim=10mm 50mm 0mm 50mm,angle=0,scale=0.35]{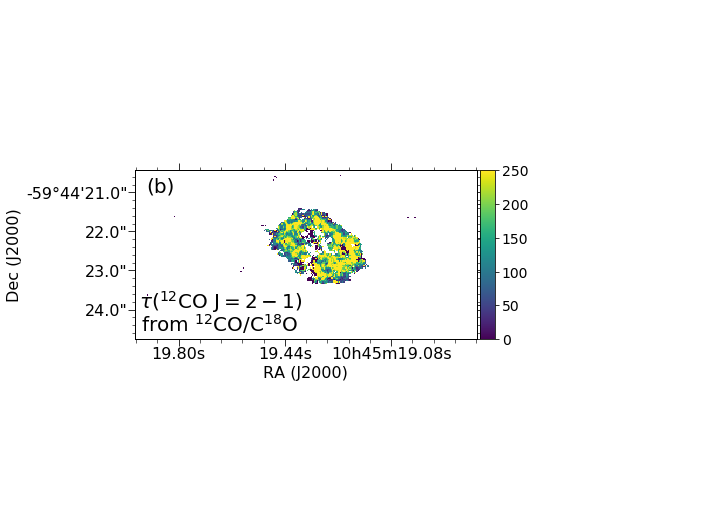} \\
    \includegraphics[trim=10mm 50mm 0mm 50mm,angle=0,scale=0.35]{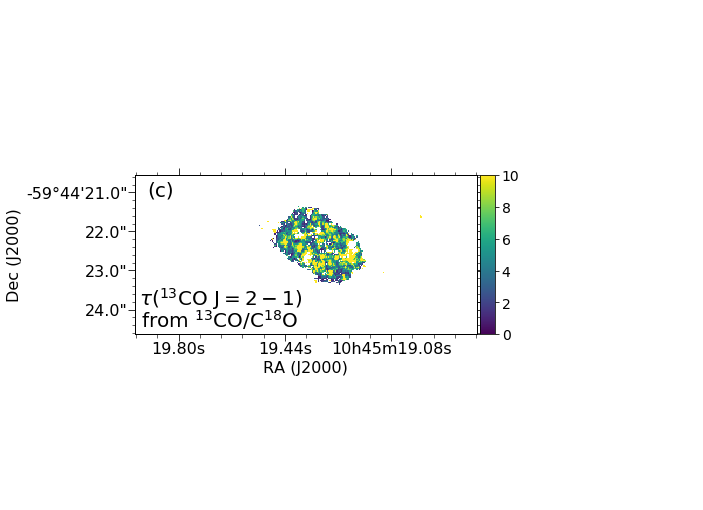} \\
    \includegraphics[trim=10mm 50mm 0mm 50mm,angle=0,scale=0.35]{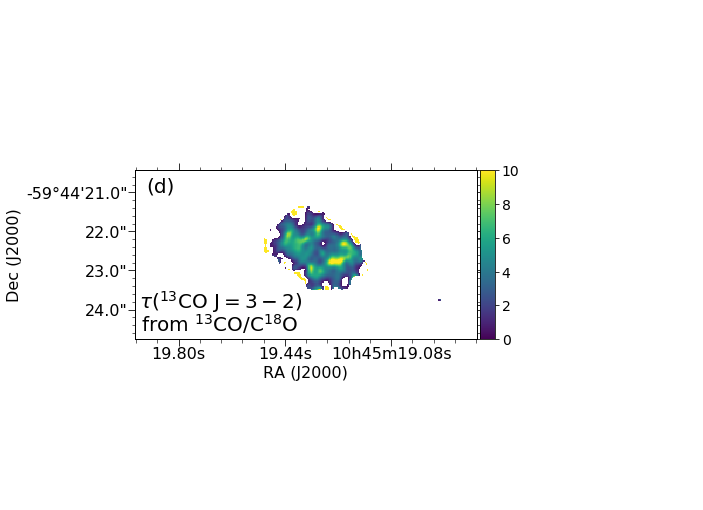} \\
    \end{array}$
\caption{Maps of the optical depth, $\tau_v$, of the (a) $^{12}$CO J=2-1
from the temperature ratio $T_{12}/T_{13}$, 
(b) $^{12}$CO J=2-1
from the temperature ratio $T_{12}/T_{18}$, 
(c) $^{13}$CO J=2-1, and (d) $^{13}$CO J=3-2. We compute $\tau$ for each pixel as a function of velocity; maps shown here display the value of $\tau_v$ at the $v_{\mathrm{LSR}} = -33.5$~\kms.} \label{fig:tau_maps}
\end{figure}

As described in Section~\ref{ss:tau}, we compute the optical depth at each position and at each velocity. 
We show maps of $\tau$ at the $v_{\mathrm{LSR}}$ in Figure~\ref{fig:tau_maps} to give an overview of the optical depth structure in the source. 
Data used in each calculation were masked for significance using the following cuts: 
$6\sigma$ for $^{12}$CO, 
$4\sigma$ for \tco, and 
$3\sigma$ for \cdo. 
Maps shown reflect a single channel. As a result, not every pixel meets the significance threshold leading to white space in a few places in the maps shown in Figure~\ref{fig:tau_maps}. 

\section{Temperature maps}\label{s:Tmaps}

\begin{figure}
  \centering%
  $\begin{array}{c}
    \includegraphics[trim=10mm 0mm 0mm 50mm,angle=0,scale=0.35]{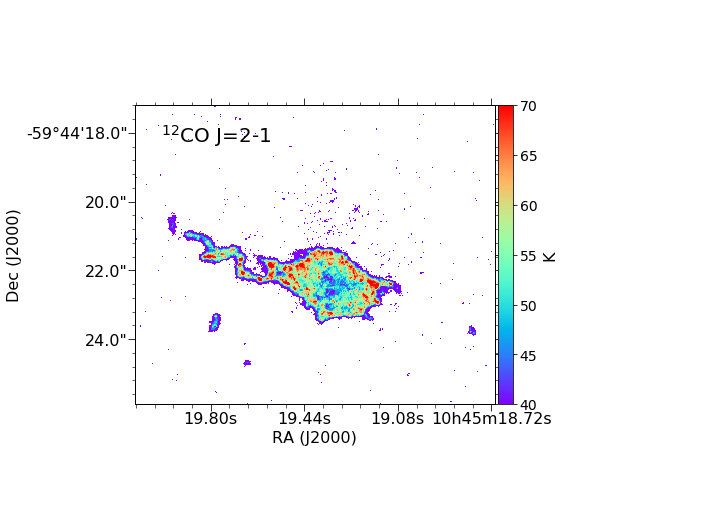} \\
    \includegraphics[trim=10mm 0mm 0mm 50mm,angle=0,scale=0.35]{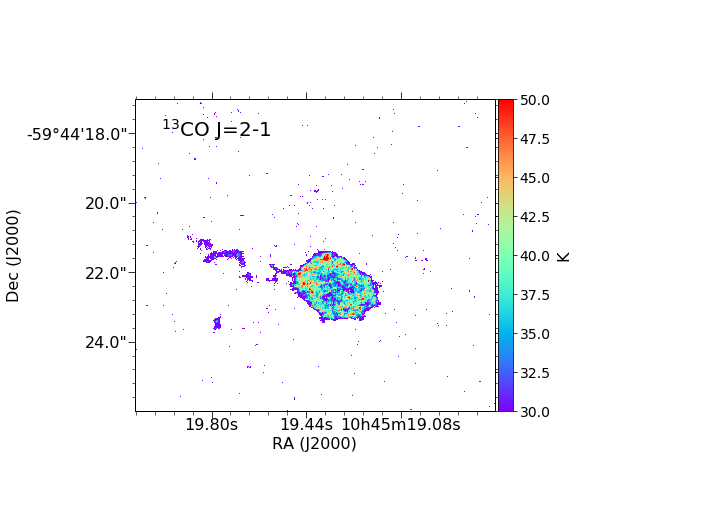} \\
    \includegraphics[trim=10mm 0mm 0mm 50mm,angle=0,scale=0.35]{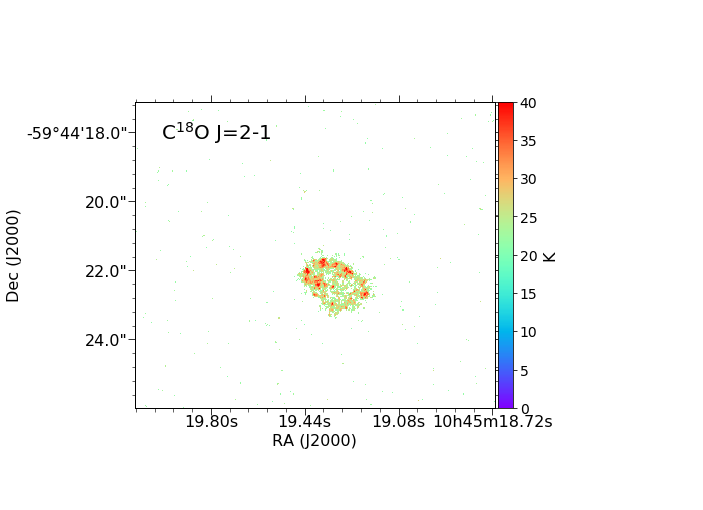} \\
    \includegraphics[trim=10mm 0mm 0mm 50mm,angle=0,scale=0.35]{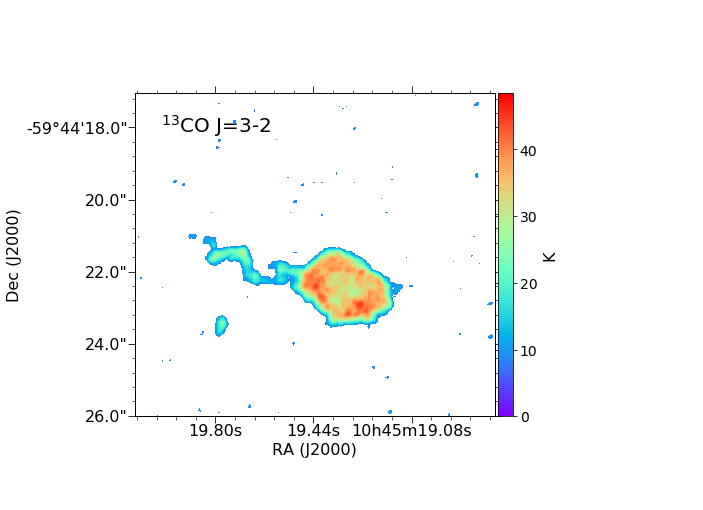} \\
    \includegraphics[trim=10mm 30mm 0mm 50mm,angle=0,scale=0.35]{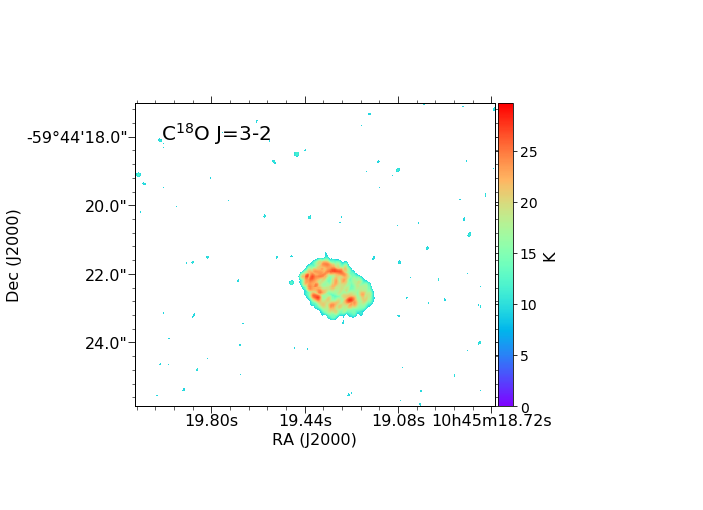} \\
    \end{array}$
\caption{Maps of the peak brightness temperature of all observed CO isotopologues. \label{fig:Tmaps}}
\end{figure}

We show a map of the peak brightness temperature of all of the observed CO isotopologues in Figure~\ref{fig:Tmaps}. 
The brightness temperature depends on the excitation temperature as $T_{mb} = T_{ex} (1-e^{-\tau})$. 
For optically thick lines ($\tau >>1$) $T_{mb} \approx T_{ex}$, allowing the brightness temperature to be taken as a proxy for the excitation temperature (see also the discussion in Section~\ref{s:tadpole_temp}). 
Assuming that all of the observed CO isotopologues are optically thick (see Section~\ref{s:tau_maps} and Figure~\ref{fig:tau_maps}), the peak brightness temperatures shown in Figure~\ref{fig:Tmaps} reflect the excitation temperature of that line at the $\tau=1$ surface.

The $\tau=1$ surface traced by rarer isotopologues and higher excitation transitions should probe deeper in the cloud than the most abundant isotopologue (and most optically thick line), $^{12}$CO J=2-1 (see Figure~\ref{fig:bTemp}).
While our data do not allow us to constrain the location of the $\tau=1$ surface, we note that the brightness temperature is lower for rarer isotopologues and higher excitation transitions. 

\section{Column density maps}\label{s:Ncol_maps}

\begin{figure*}
  \centering%
$\begin{array}{ccc}
\includegraphics[trim=30mm 50mm 20mm 30mm,angle=0,scale=0.265]{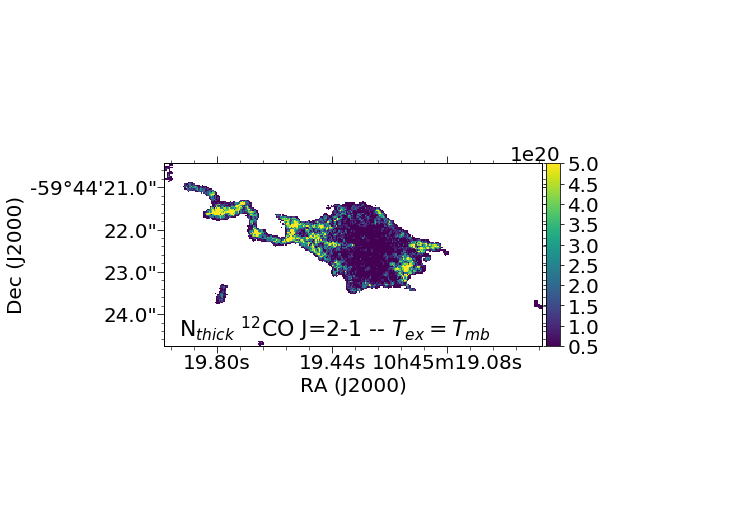} &
\includegraphics[trim=30mm 50mm 20mm 30mm,angle=0,scale=0.265]{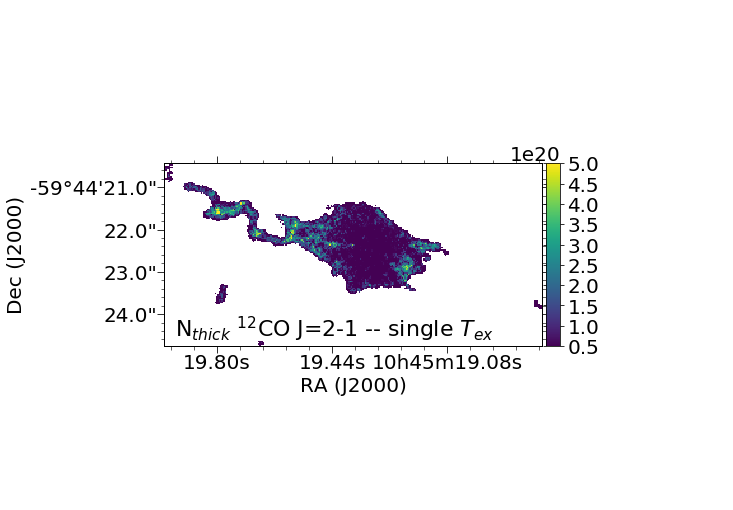} &
\includegraphics[trim=30mm 50mm 20mm 30mm,angle=0,scale=0.265]{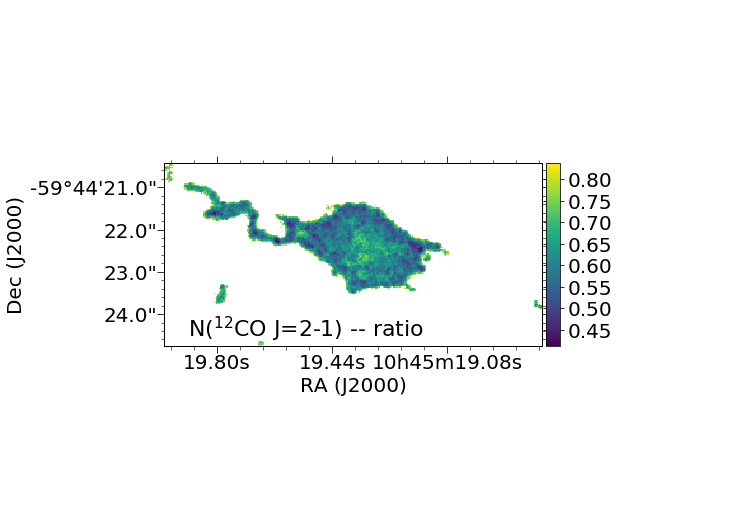} \\
\includegraphics[trim=30mm 50mm 20mm 30mm,angle=0,scale=0.265]{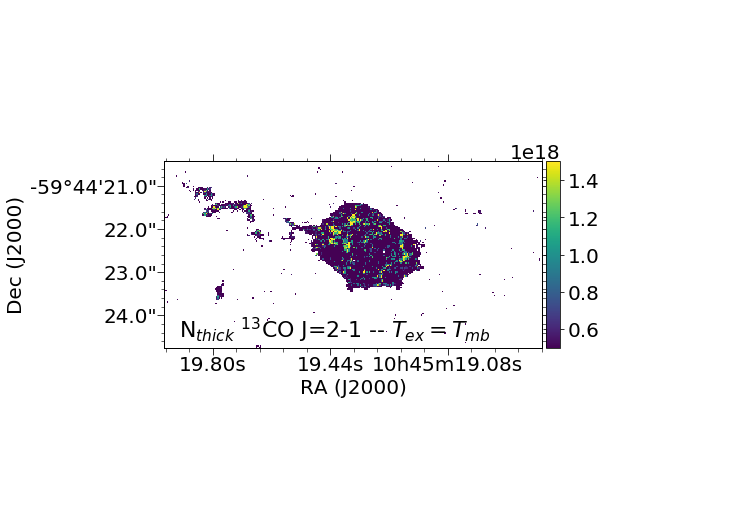} &
\includegraphics[trim=30mm 50mm 20mm 30mm,angle=0,scale=0.265]{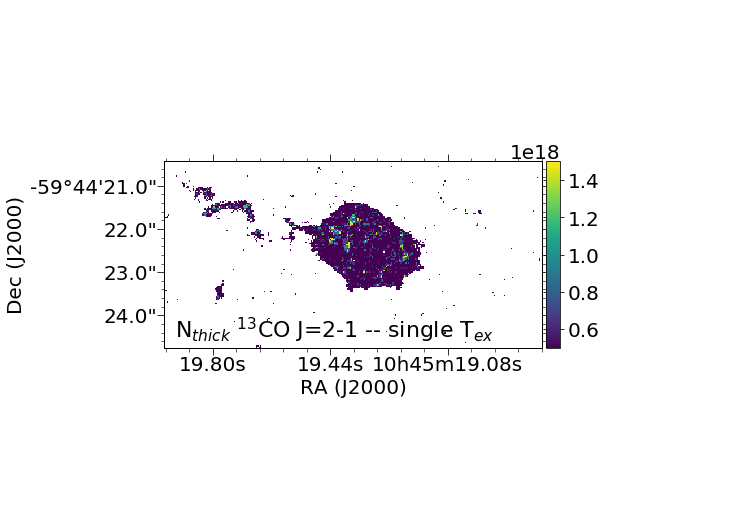} &
\includegraphics[trim=30mm 50mm 20mm 30mm,angle=0,scale=0.265]{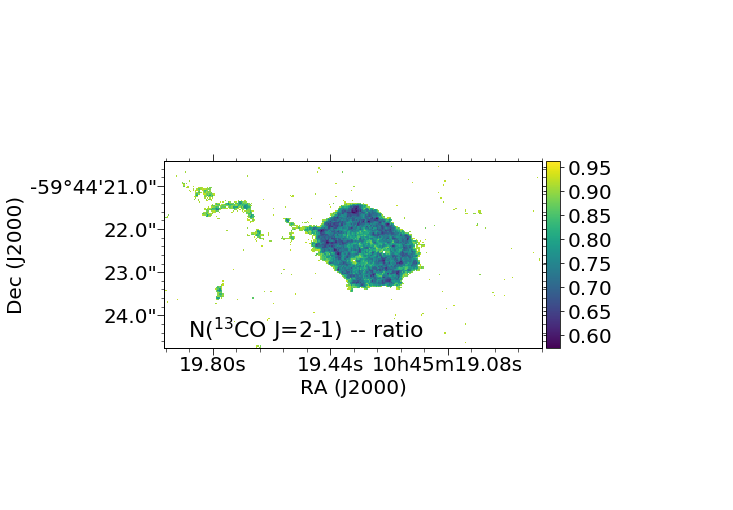} \\
\includegraphics[trim=30mm 50mm 20mm 30mm,angle=0,scale=0.265]{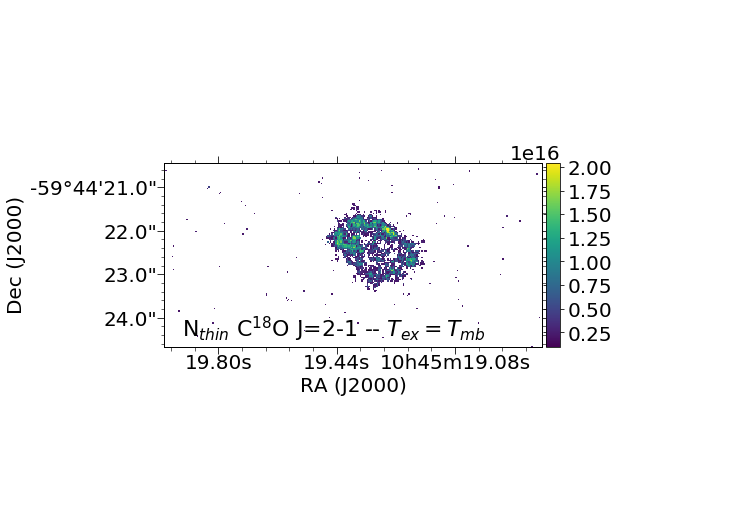} &
\includegraphics[trim=30mm 50mm 20mm 30mm,angle=0,scale=0.265]{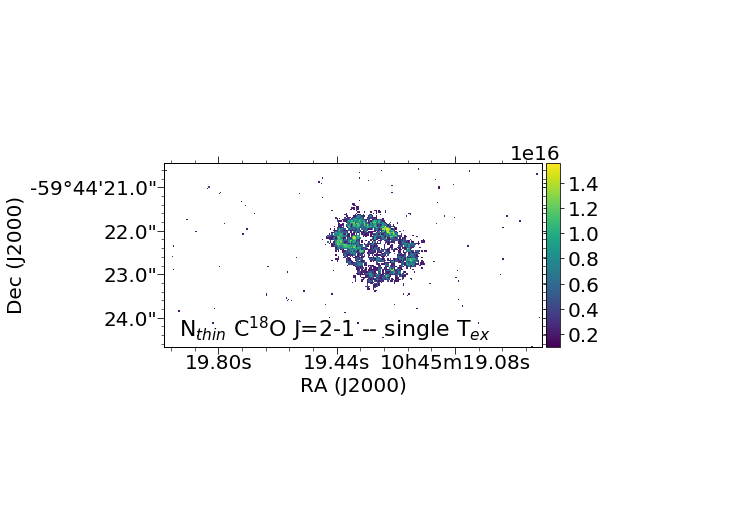} &
\includegraphics[trim=30mm 50mm 20mm 30mm,angle=0,scale=0.265]{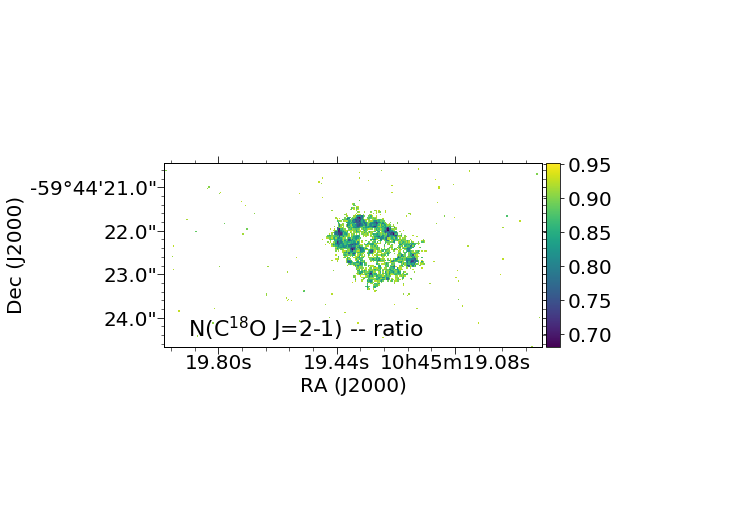} \\
\includegraphics[trim=30mm 50mm 20mm 30mm,angle=0,scale=0.265]{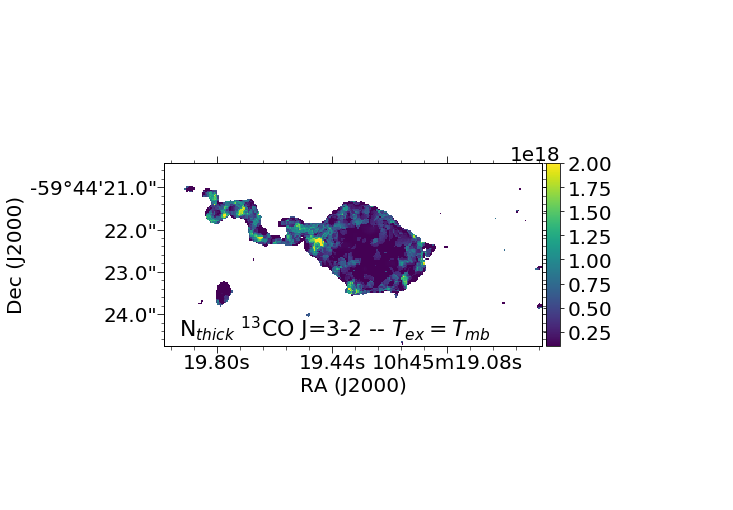} &
\includegraphics[trim=30mm 50mm 20mm 30mm,angle=0,scale=0.265]{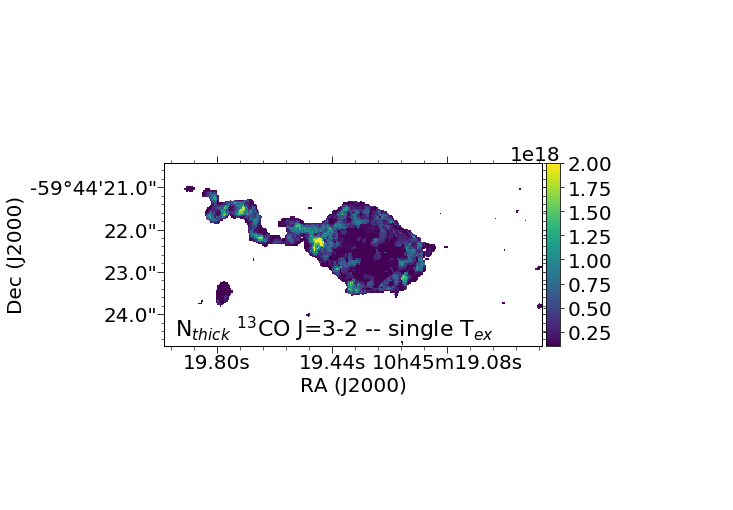} &
\includegraphics[trim=30mm 50mm 20mm 30mm,angle=0,scale=0.265]{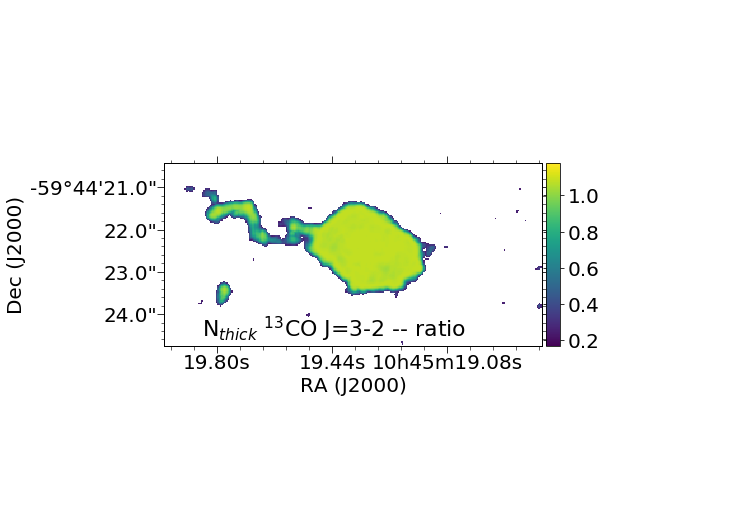} \\
\includegraphics[trim=30mm 50mm 20mm 30mm,angle=0,scale=0.265]{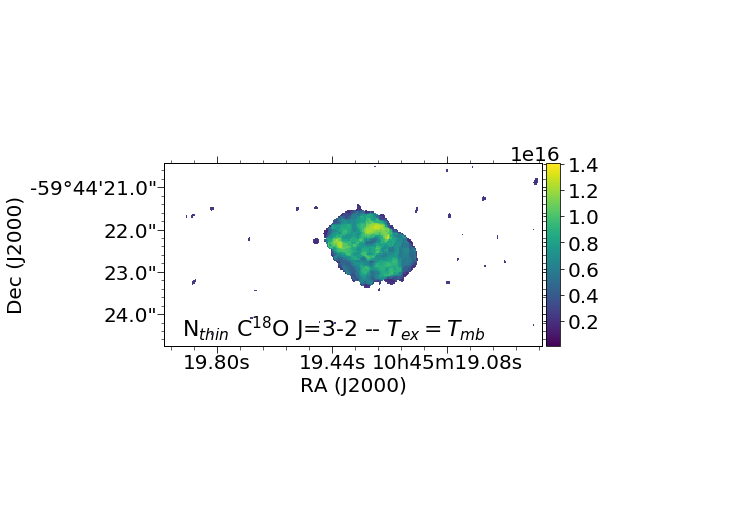} &
\includegraphics[trim=30mm 50mm 20mm 30mm,angle=0,scale=0.265]{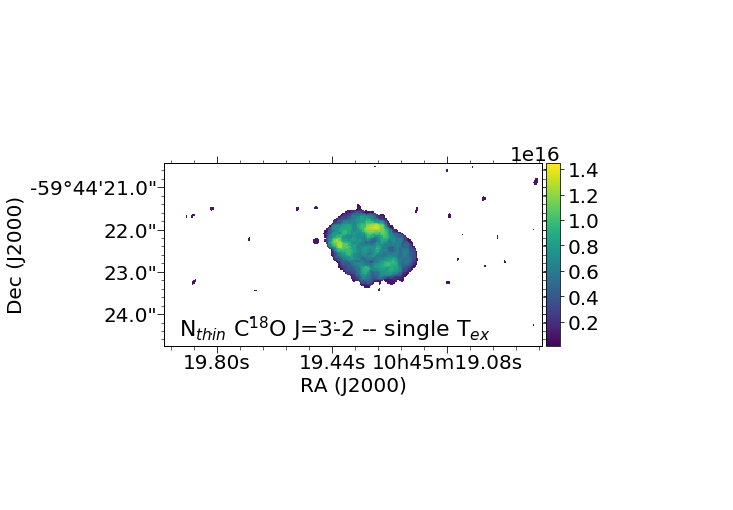} &
\includegraphics[trim=30mm 50mm 20mm 30mm,angle=0,scale=0.265]{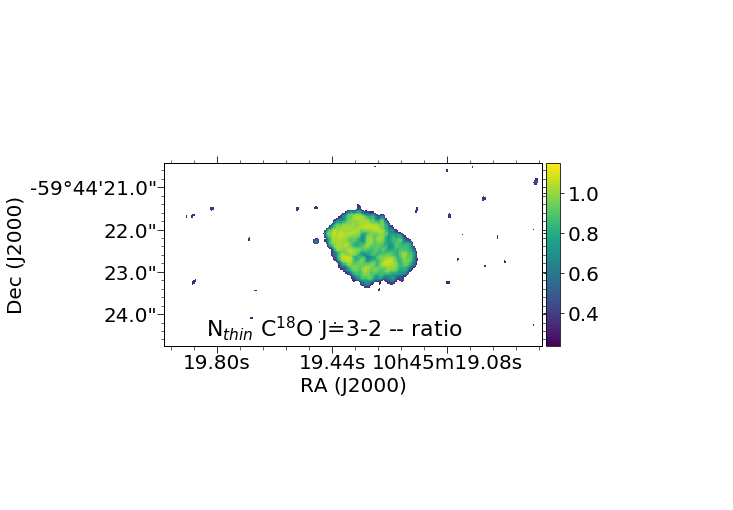} \\
\end{array}$
\caption{Maps of the column density of all observed transitions of CO and its isotoptologues. In the left column, we show the column density computed using the brightness temperature (see Section~\ref{s:Tmaps} and Figure~\ref{fig:Tmaps}) as a proxy for excitation temperature (see Section~\ref{s:tadpole_temp}).  
The center column shows the column density assuming a single excitation temperature, $T_{ex}=20$~K, as in Section~\ref{ss:alma_NH}. The right column shows the ratio of the two column density estimates (single $T_{ex}$ / variable $T_{ex}$). 
 \label{fig:CO_Ncol_maps}}
\end{figure*}

We show maps of the column density of the CO isotopologues in Figure~\ref{fig:CO_Ncol_maps}. 
In Section~\ref{ss:alma_NH}, we compute the column density assuming a single excitation, $T_{ex}=20$~K and argue that higher excitation temperatures will change this estimate by factors of $\sim 2$. To confirm this, we recompute the column density of the CO lines assuming that the brightness temperature traces the excitation (i.e.\ using the temperature structure show in Figure~\ref{fig:Tmaps}). To compute the appropriate $Q(T_{ex})$ at each position, we take the discrete values of $Q(T_{ex})$ for each isotopologue given in the JPL Spectral Line Catalog \citep{pickett1998} and fit a third-order polynomial. 
Column densities computed this way are shown in the left column of Figure~\ref{fig:CO_Ncol_maps}. 
We compare this with the column densities we estimate assuming a single $T_{ex}$ (shown in the middle column of Figure~\ref{fig:CO_Ncol_maps}). 
The ratio of the two column density calculations are shown in the right column of Figure~\ref{fig:CO_Ncol_maps}; values typically vary by a factor of $\lesssim 2$.

%

\bsp	
\label{lastpage}
\end{document}